\documentclass[twocolumn]{aastex631}
\usepackage{amsmath}
\usepackage{graphicx}
\usepackage{soul}

\received{Oct. 20, 2021}
\revised{Nov. 25, 2021}
\accepted{Nov. 29, 2021}

\submitjournal{ApJ}

\shorttitle{Love numbers of rapid rotators}
\shortauthors{Dewberry \& Lai}
\graphicspath{{./}{figures/}}

\begin{document}

\title{Dynamical tidal Love numbers of rapidly rotating planets and stars}

\correspondingauthor{Janosz W. Dewberry}
\email{dewberry@caltech.edu}

\author[0000-0001-9420-5194]{Janosz W. Dewberry}
\affiliation{Canadian Institute for Theoretical Astrophysics, 60 St. George Street, Toronto, ON M5S 3H8, Canada}
\affiliation{Division of Physics, Mathematics and Astronomy, California Institute of Technology, Pasadena, CA 91125, USA}

\author[0000-0002-1934-6250]{Dong Lai}
\affiliation{Cornell Center for Astrophysics and Planetary Science, Department of Astronomy, Cornell University, Ithaca, NY 14853, USA}



\begin{abstract}
Tidal interactions play an important role in many astrophysical systems, but uncertainties regarding the tides of rapidly rotating, centrifugally distorted stars and gaseous planets remain. We have developed a precise method for computing the dynamical, non-dissipative tidal response of rotating planets and stars, based on summation over contributions from normal modes driven by the tidal potential. We calculate the normal modes of isentropic polytropes rotating at up to $\simeq90\%$ of their critical breakup rotation rates, and tabulate fits to mode frequencies and tidal overlap coefficients that can be used to compute the frequency-dependent, non-dissipative tidal response (via potential Love numbers $k_{\ell m}$). Although fundamental modes (f-modes) possess dominant tidal overlap coefficients at (nearly) all rotation rates, we find that the strong coupling of retrograde inertial modes (i-modes) to tesseral ($\ell>|m|$) components of the tidal potential produces resonances that may be relevant to gas giants like Jupiter and Saturn. The coupling of f-modes in rapid rotators to multiple components of both the driving tidal potential and the induced gravitational field also affect the tesseral response, leading to significant deviations from treatments of rotation that neglect centrifugal distortion and high-order corrections. For very rapid rotation rates ($\gtrsim 70\%$ of breakup), mixing between prograde f-modes and i-modes significantly enhances the sectoral ($\ell=|m|$) tidal overlap of the latter. The tidal response of very rapidly rotating, centrifugally distorted planets or stars can also be modified by resonant sectoral f-modes that are secularly unstable via the Chandrasekhar-Friedman-Schutz (CFS) mechanism.
\end{abstract}



\section{Introduction}\label{sec:intro}
The question of how a self-gravitating fluid body (such as a star or a planet) responds to the gravitational potential of an orbiting satellite becomes complicated, especially at a quantitative level, when the body in question rotates at a significant fraction of the ``breakup'' rotation rate, $\Omega_d=(GM/R_\text{eq}^3)^{1/2}$ (where $M$ and $R_\text{eq}$ are the body's mass and equatorial radius). Recent measurements of the non-spherical gravity fields of Jupiter and Saturn provided by \emph{Juno} and \emph{Cassini} promise insight into this problem, since the two planets possess relatively rapid bulk rotation rates of $\Omega\simeq0.3\Omega_d$ and $\simeq0.4\Omega_d$ (respectively). The satellite data have facilitated the placement of observational constraints on the (dissipative as well as non-dissipative) tidal response for both planets \citep{Lainey2009,Lainey2012,Lainey2017,Lainey2020,Durante2020}. In particular, gravity field measurements allow for the inference of the so-called `potential Love numbers' $k_{\ell m}$---response functions that measure the ratios between the harmonic components of the gravitational potentials associated with tidal deformation of the planet, and components of the potentials imposed by planetary satellites \citep[e.g.,][]{Ogilvie2014}. These unprecedented measurements have the potential to constrain planetary internal structures, and invite further theoretical investigations into the dynamical tides of rapidly rotating planets and stars.

Previous efforts in the planetary sciences community have primarily adopted 3D `Concentric MacLaurin Spheroid' (CMS) calculations of distorted planetary structure \citep{Wahl2017,Wahl2020,Nettelmann2019}. These calculations rely on a hydrostatic approximation, neglecting the finite tidal forcing frequencies associated with Jupiter and Saturn's moons. Significant discrepancies between the observed Love numbers and those calculated with the CMS method for Jupiter indicate that dynamical, non-hydrostatic effects modify the planet's tidal response. 

\citet{Idini2021} showed through direct computations incorporating the Coriolis force that dynamical tides can explain the Love number discrepancies, in particular for the quadrupolar $k_{22}$. \citet{Lai2021} demonstrated that the same results can be achieved more efficiently with a phase space expansion in the normal mode oscillations of the planet. However, \citet{Lai2021} employed a perturbative treatment of rotation (to linear order in $\Omega$), and both \citet{Idini2021} and \citet{Lai2021} ignored the $\sim\mathcal{O}(\Omega^2$) effects of the centrifugal distortion of the equilibrium state on the fundamental `f-modes' that typically dominate the gravitational response. Jupiter's rapid rotation makes the accuracy of perturbative treatments unclear. 

In this paper, we use a non-perturbative method to compute the normal modes of rapidly rotating fluid bodies. Focusing on $n=1$ polytropes with rotation rates $\Omega\lesssim0.9\Omega_d$, we tabulate fits to the mode properties required to calculate the (frequency-dependent) potential Love numbers $k_{\ell m}$ with azimuthal wave numbers $m=1,2,3,4$ and spherical harmonic degrees $\ell=2,3,4,5,6$ (for even $\ell-m$). In addition to giant planets, whose structures can be well approximated by $n=1$ polytropes, our results can also be applied approximately to rapidly rotating neutron stars. 

As in the non-rotating limit, f-modes dominate the non-dissipative tidal response at all rotation rates, except near resonances where the tidal forcing frequency matches the frequency of a mode. In particular, inertial modes (i-modes) restored by the Coriolis force have frequencies that can resonate with the tidal forcing. The i-modes also possess a strong tesseral ($\ell>|m|$) overlap that is often not considered in studies of tidal interactions focused on, e.g., the quadrupolar component of the perturbing potential \citep[but see, e.g., ][]{Ogilvie2013,Braviner2015}. This tesseral coupling may be relevant to discrepant observed and predicted values of tesseral Love numbers \citep[such as $k_{42}$; ][]{Durante2020}. In addition, our non-perturbative inclusion of f-modes' overlap with multiple spherical harmonics due to centrifugal distortion
also leads to significant (non-resonant) deviations of the tesseral Love numbers $k_{m+2,m}$ from the calculations of \citet{Lai2021}.

This paper is structured as follows. In Section \ref{sec:methods} we describe our methods for calculating centrifugally distorted stellar models, normal mode oscillations and potential Love numbers. In Section \ref{sec:modes} we review the general properties of the oscillation modes considered. We present results in Section \ref{sec:fires}, and conclude in Section \ref{sec:conc}.

\section{Methods}\label{sec:methods}
In this section we outline our methods for calculating oblate models of rotating polytropes (Section \ref{sec:modelcalc}), and their normal mode oscillations (Section \ref{sec:modecalc}). We then introduce the relevant quantities required to characterize the response of a fluid body to an imposed tidal potential (Section \ref{sec:tides}). Throughout the paper, we refer both to spherical and cylindrical polar coordinates denoted by $(r,\theta,\phi)$ and $(R,\phi,z)$, respectively.

\subsection{Stellar and planetary models}\label{sec:modelcalc}
We consider rapidly rotating, neutrally stratified (i.e., fully and efficiently convective) polytropes characterized by the barotropic equation of state $P\propto\rho^{1+1/n}$, focusing on the polytropic index $n=1$. The convective envelopes in current models of Jupiter and Saturn match relatively well with $n=1$ polytropes, although they may differ both in details and in the deep interior \citep[e.g.,][]{Mankovich2021}. Newtonian polytropes with $n=1$ are also frequently used to approximate the interiors of neutron stars \citep[e.g.,][]{Passamonti2009a,Xu2017}, since this index reproduces predictions by equations of state for nuclear matter of an approximately constant neutron star radius over a range of masses around $1.4M_\odot$ \citep[see, e.g.,][]{Lattimer2007}. The inclusion of super-fluidity \citep{Passamonti2009b} and general relativity \citep{dePietri2018} are, however, required for more realistic representations.

We use a two-dimensional pseudospectral method to calculate axisymmetric but oblate polytropic models. It is reasonable to start from an axisymmetric basic state when computing the \emph{linear} tidal response; the tidal deformations of Jupiter and Saturn, for instance, are very weak in comparison with their rotational deformations. \autoref{app:modelcalc} provides the details of these calculations. In brief, we start by calculating one-dimensional, non-rotating solutions to the Lane-Emden equation, and then use an iterative scheme to compute centrifugally distorted models. We verify the accuracy of our model calculations by considering virial errors.

In addition to the polytropic index $n$ and rotation rate $\Omega$, different models are characterized by the equatorial and polar radii $R_\text{eq}$ and $R_\text{pol}$ (resp.), the central value for the pseudo-enthalpy $H=\int \rho^{-1}\text{d}P=(1+n)P/\rho,$ the total mass, and the ratio between kinetic and potential energy. Unless otherwise stated, we use units with $G=M=R_\text{eq}=1$ so that angular velocities take on units of $\Omega_d=(GM/R_\text{eq}^3)^{1/2}$ ($\Omega_d$ is frequently referred to as the ``dynamical frequency''). 

\subsection{Mode calculations}\label{sec:modecalc}
In the limit of zero rotation, the partial differential equations (PDEs) governing linear perturbations to a self-gravitating fluid in hydrostatic equilibrium are separable in terms of spherical harmonics. Rotation disrupts this separability, since both the Coriolis force and centrifugal distortion of the equilibrium state break spherical symmetry. Perturbative treatments, in which eigenfunctions and frequencies are expanded in powers of $\Omega$, prove adequate when rotation is slow \citep[see, e.g.,][]{Unno1989}. However, the normal mode oscillations of fluid bodies rotating at an appreciable fraction of the equatorial Keplerian frequency $\Omega_d$ require more complete treatments, as do modes with frequencies comparable to the rotation rate (even when $\Omega\ll\Omega_d$).

Non-perturbative methods \citep[e.g.,][]{Lignieres2006,Reese2006,Reese2009,Reese2013,Reese2021,Ouazzani2012} treat the linearized PDEs as fundamentally non-separable, eschewing assumptions of spherical symmetry and instead solving an inherently two-dimensional problem. Having calculated equilibrium pressure and density profiles $P_0(r,\theta)$ and $\rho_0(r,\theta)$ for a given polytropic index $n$ and rotation rate $\Omega$, we use the spectral method described in \citet{Dewberry2021} to solve for adiabatic normal modes with the time dependence $\exp[-\text{i}\omega t]$ in the rotating frame. We refer interested readers to that work for the details of these calculations, but make note of a few points salient to this investigation:

In a rapidly rotating planet or star, normal mode eigenfunctions involve Eulerian perturbations ${\bf v},\rho',P',\Phi'$ to the fluid velocity, density, pressure and gravitational field (resp.) that depend non-trivially on radius and polar angle; given the non-separability of the governing PDEs, mode eigenfunctions cannot be associated with a single spherical harmonic $Y_\ell^m$. Instead, the non-perturbative approach involves a more agnostic expansion in series of spherical harmonics. For example, we represent the gravitational potential perturbation $\Phi'(r,\theta,\phi,t)$ with an expansion of the form
\begin{equation}\label{eq:PhiExp}
    \Phi'
    =\mathcal{R}e\left\{
        \sum_{\ell'=m}^{\ell_\text{max}}\Phi^{\ell'}(\zeta)
        Y_{\ell'}^m(\theta,\phi)
        \exp[-\text{i}\omega t]
    \right\}.
\end{equation}
Note that because the equilibrium states are axisymmetric, the equations are still separable in $\phi$, and each oscillation mode can be identified with a unique azimuthal wavenumber $m.$ We adopt the convention of strictly positive $m$, with prograde (retrograde) propagation in the rotating frame differentiated by positive (negative) $\omega$. The coefficients $\Phi^\ell(\zeta)$ in this expansion depend on a quasi-radial coordinate $\zeta$ associated with a non-orthogonal coordinate system originally introduced by \citet{Bonazzola1998}, which is constructed to match the oblate surface $r_s(\theta)$ of the rotating fluid body (at $\zeta/R_\text{eq}=1$), and to relax to spherical coordinates both at the origin ($\zeta=0$) and at the edge of the exterior vacuum region included in the computational domain (defined by $\zeta/R_\text{eq}=2$). 

We solve Laplace's equation $\nabla^2\Phi'=0$ in the external vacuum, which we include for the purposes of boundary condition application: at the oblate surface $r_s$ we impose the continuity of $\Phi'$ and its gradient, in addition to the free surface condition of a vanishing Lagrangian pressure perturbation. Following \citet{Reese2006}, we apply the actual boundary condition for the gravitational potential on the spherical surface $\zeta=r=2R_\text{eq}$, matching harmonic coefficients $\Phi^\ell$ to the analytical solutions that vanish at infinity. We enforce the standard regularity conditions at the origin \citep[see Chapter III in ][]{Unno1989} implicitly.

\subsection{Tides and Love numbers}\label{sec:tides}
We work in a frame centred on, and rotating with the tidally disturbed planet or star (hereafter ``planet'') under consideration. In the vicinity of this planet, a satellite with mass $M'$ and orbital separation $a$ on a circular and co-planar orbit produces a gravitational potential that can be expanded as
\begin{equation}\label{eq:tpot}
    U=\sum_{\ell,m}
    U_{\ell m}r^\ell Y_\ell^m(\theta,\phi)
    \exp[-\text{i}\omega_mt],
\end{equation}
where the tidal forcing frequency is
\begin{equation}
    \omega_m=m(\Omega_o-\Omega),
\end{equation}
and $\Omega_o=[G(M+M')/a^3]^{1/2}$ is the orbital frequency. The harmonic coefficients $U_{\ell m}$ are then given by \citep{Jackson1962,Press1977} 
\begin{equation}
    U_{\ell m}=-\frac{GM'}{a^{\ell+1}}W_{\ell m},
\end{equation}
where $W_{\ell m}=0$ for non-integer $(\ell+m)/2$, and otherwise
\begin{align}
    W_{\ell m}
    &=\left(\frac{4\pi}{2\ell + 1}\right)Y_\ell^{m*}(\pi/2,0)
\\\notag
    &=(-1)^{(\ell+m)/2}
    \left[
        \frac{4\pi(\ell+m)!(\ell-m)!}{(2\ell + 1)}
    \right]
    \\\notag&
    \hspace{3.75em}
    \times
    \left[
        2^\ell
        \left( 
            \frac{\ell+m}{2}
        \right)! 
        \left( 
            \frac{\ell-m}{2}
        \right)!
    \right]^{-1}.
\end{align}

The tidal potential $U$ drives density perturbations in the planet, which then generate an external potential perturbation that can be expanded as
\begin{equation}\label{eq:exm}
    \delta\Phi=\sum_{\ell,m}
    \delta\Phi_{\ell m}r^{-(\ell+1)}
    Y_\ell^m\exp[-\text{i}\omega_mt].
\end{equation}
The potential Love number $k_{\ell m}$ is then defined as the ratio \citep[e.g.,][]{Ogilvie2014}
\begin{equation}\label{eq:klm}
    k_{\ell m}=\frac{\delta\Phi_{\ell m}}{U_{\ell m}}.
\end{equation}
In the absence of any dissipative processes, $k_{\ell m}$ is real-valued; we ignore the Love numbers' imaginary parts, but note that they can be significant near resonances, or in rotating planets with solid cores.

In the linear theory, the tidal response of the planet can be written as an expansion in normal modes \citep[e.g.,][]{Schenk2002,Lai2006}. Specifically, we write
\begin{equation}
    k_{\ell m}=\sum_\alpha k_{\ell m}^\alpha,
\end{equation}
where $k_{\ell m}^\alpha$ describes the contribution of a mode labelled by $\alpha$. Following the procedure of \citep{Lai2021}, with the inclusion of higher-order centrifugal effects we find
\begin{equation}\label{eq:klma}
    k_{\ell m}^\alpha
    =\frac{2\pi}{(2\ell+1)}
    \sum_{\ell'}
    \frac{Q_{\ell m}^\alpha Q_{\ell' m}^\alpha}
    {\epsilon_{\alpha}(\omega_\alpha-\omega_m)}
    \left(\frac{ U_{\ell' m}}{ U_{\ell m}}\right).
\end{equation}
Defining the inner product 
$\langle \boldsymbol{\xi},\boldsymbol{\xi}\rangle
=\int_V\rho_0(r,\theta)\boldsymbol{\xi}^*\cdot \boldsymbol{\xi}\text{d}V$ for  the Lagrangian displacement $\boldsymbol{\xi}=(-\text{i}\omega)^{-1}{\bf v}$, \autoref{eq:klma} involves the quantity 
\begin{equation}\label{eq:eps}
    \epsilon_\alpha
    =\omega_\alpha
    \langle \boldsymbol{\xi}_\alpha,\boldsymbol{\xi}_\alpha\rangle
    +\langle\boldsymbol{\xi}_\alpha,
    \text{i}{\bf \Omega}\times\boldsymbol{\xi}_\alpha\rangle,
\end{equation}
along with the tidal overlap coefficient
\begin{align}\label{eq:Qlm}
    Q_{\ell m}^\alpha
    &=
    \langle \boldsymbol{\xi}_\alpha,\nabla (r^\ell Y_\ell^m)\rangle
\\\notag
    &=
    \int_V r^\ell Y_\ell^m\rho'^*_\alpha\text{d} V
\\\notag
    &=-\frac{(2\ell+1)}{4\pi}
    \left[
        2^{\ell+1}
        \Phi_\alpha^\ell(\zeta)|_{\zeta=2}
    \right].
\end{align}
Note that all quantities in \autoref{eq:klma} are dimensionless, under our adoption of units with $G=M=R_\text{eq}=1.$ 

Each coefficient $Q_{\ell m}^\alpha$ characterizes the spatial overlap of the mode $\alpha$ with the $\ell m$ component of the tidal potential. In the last equality of \autoref{eq:Qlm}, the quantity in square brackets gives the coefficient of degree $\ell$ in a harmonic expansion \emph{in spherical coordinates} of the mode's gravitational potential perturbation (at radius $r=R_\text{eq}$). This can differ significantly from the coefficient $\Phi^\ell_\alpha(\zeta)|_{\zeta=1}$ in the expansion of \autoref{eq:PhiExp}, since the rotating body is non-spherical. 

\autoref{eq:klma} is more general than Eq. 11 in \citet{Lai2021}, in that it includes an additional sum over spherical harmonic degrees indexed by $\ell'$. This sum is necessary in order to fully account for $\sim\mathcal{O}(\Omega^2)$ centrifugal effects on the planet's structure and modes. In a spherically symmetric planet, all terms with $\ell'\not=\ell$ would vanish, and the ratio $U_{\ell'm}/U_{\ell m}$ would collapse to one. More generally,
\begin{equation}\label{eq:Uratio}
    \frac{U_{\ell'm}}{U_{\ell m}}
    =\frac{W_{\ell' m}}{W_{\ell m}}
    \frac{a^\ell}{a^{\ell'}}
    =\frac{W_{\ell' m}}{W_{\ell m}}
    \left[
        \frac{(1+q)}{\Omega_o^2}
    \right]^{(\ell-\ell')/3}
    ,
\end{equation}
where $q=M'/M$ is the mass ratio [recall that $a^{\ell-\ell'}=(a/R_\text{eq})^{\ell-\ell'}$ in our units]. 

Thus $U_{\ell'm}/U_{\ell m}\propto \Omega_o^{-2(\ell -\ell')/3}$, which increases as the separation $a$ increases ($\Omega_o$ decreases) for $\ell-\ell'>0$. This leads to a divergence in some $k_{\ell m}$ (specifically tesseral Love numbers with $\ell>m$) as $a\rightarrow\infty$. For example, consider $k_{42}$ ($\ell=4,m=2$): in a centrifugally flattened planet, modes excited by the quadrupolar $U_{22}$ potential contribute to $\delta\Phi_{42}$. As $a\rightarrow\infty$, the quadrupolar contribution to $\delta\Phi_{42}$ can in fact dominate the contribution from $U_{42},$ simply because $|U_{22}|\gg |U_{42}|$. The ratio $k_{42}=\delta\Phi_{42}/U_{42}$ then ``diverges,'' even though the gravitational response $\delta\Phi_{42}$ to the satellite remains finite. The unphysical nature of this divergence, which we discuss further in Section \ref{sec:klmcalc}, suggests that Love numbers as defined by \autoref{eq:klm} to relate like-coefficients (i.e., those with the same $\ell$) in harmonic expansions may not provide the best description of the tidal response of rotationally deformed bodies.

Computing $k_{\ell m}^\alpha$ therefore requires knowledge of i) the mode frequency $\omega_\alpha$, ii) $\epsilon_\alpha$, and iii) $Q_{\ell m}^\alpha$ (for multiple $\ell$). The need for $\epsilon_\alpha$ can be eliminated by normalizing $\boldsymbol{\xi}_\alpha$ such that  $\epsilon_\alpha/\omega_\alpha=1$ \citep[see, e.g.,  ][]{Dewberry2021}, but we adopt the normalization $\langle \boldsymbol{\xi}_\alpha,\boldsymbol{\xi}_\alpha\rangle=1$ \citep[as in][]{Lai2021} because in the perturbative regime, $\epsilon_\alpha$ is simply the mode frequency in the nonrotating limit. This normalization additionally ensures that the rotational correction to the f-modes' $Q_{\ell m}^\alpha$ is of order ${\cal O}(\Omega^2)$.

\section{Mode compendium}\label{sec:modes}
In this paper we focus on the properties of fundamental modes (f-modes), inertial modes (i-modes), and to a lesser extent acoustic modes (p-modes), deferring consideration of stable stratification and internal gravity modes (g-modes) to later works. We additionally focus on modes with even equatorial parity (i.e., even $\ell-m$), which are the most relevant to planets with spin-aligned satellites. The following subsections briefly review the relevant properties of these different types of modes.

\subsection{Fundamental modes}\label{sec:fmode}
Fundamental modes (f-modes), also known as surface gravity modes, manifest even in homogeneous and incompressible stellar models. They usually possess the simplest eigenfunctions of all non-radial oscillations, with no nodes in the radial direction, and have the largest magnitude tidal overlap coefficients $Q_{\ell m}^\alpha.$ We label f-modes with the notation $f_{\ell m}^\pm$, where superscript $+/-$ denote prograde/retrograde propagation, and subscript $\ell$ is the spherical harmonic degree uniquely associated with each mode in the non-rotating limit. With increasing rotation the f-mode eigenfunctions come to involve a large number of spherical harmonic degrees, but generally maintain a clear correspondence to the non-rotating $\ell.$

\subsection{Acoustic modes}\label{sec:pmode}
Compressible ($n>0$) polytropic models also support acoustic oscillations (p-modes). Their contributions to the potential Love numbers are usually unimportant, because of their high frequencies and small $|Q_{\ell m}^\alpha|$ values in comparison with f-modes. We nonetheless include a few for completeness; employing the notation $p_{\ell m n_p}^\pm$, where $n_p$ is the number of radial nodes in the nonrotating limit, we track even-parity p-modes with $\ell<7$, $(\ell-m)/2+n_p\leq3$ for $\Omega=0$. Despite minimal frequency shifts due to the Coriolis force, strong centrifugal distortion at high rotation rates can complicate p-mode eigenfunctions and frequency spectra \citep{Lignieres2009,Reese2009,Reese2013}. However, the low-degree, low-order p-modes that we consider generally maintain a regular spectrum with increasing rotation.

\begin{figure*}
    \centering
    \includegraphics[width=\textwidth]{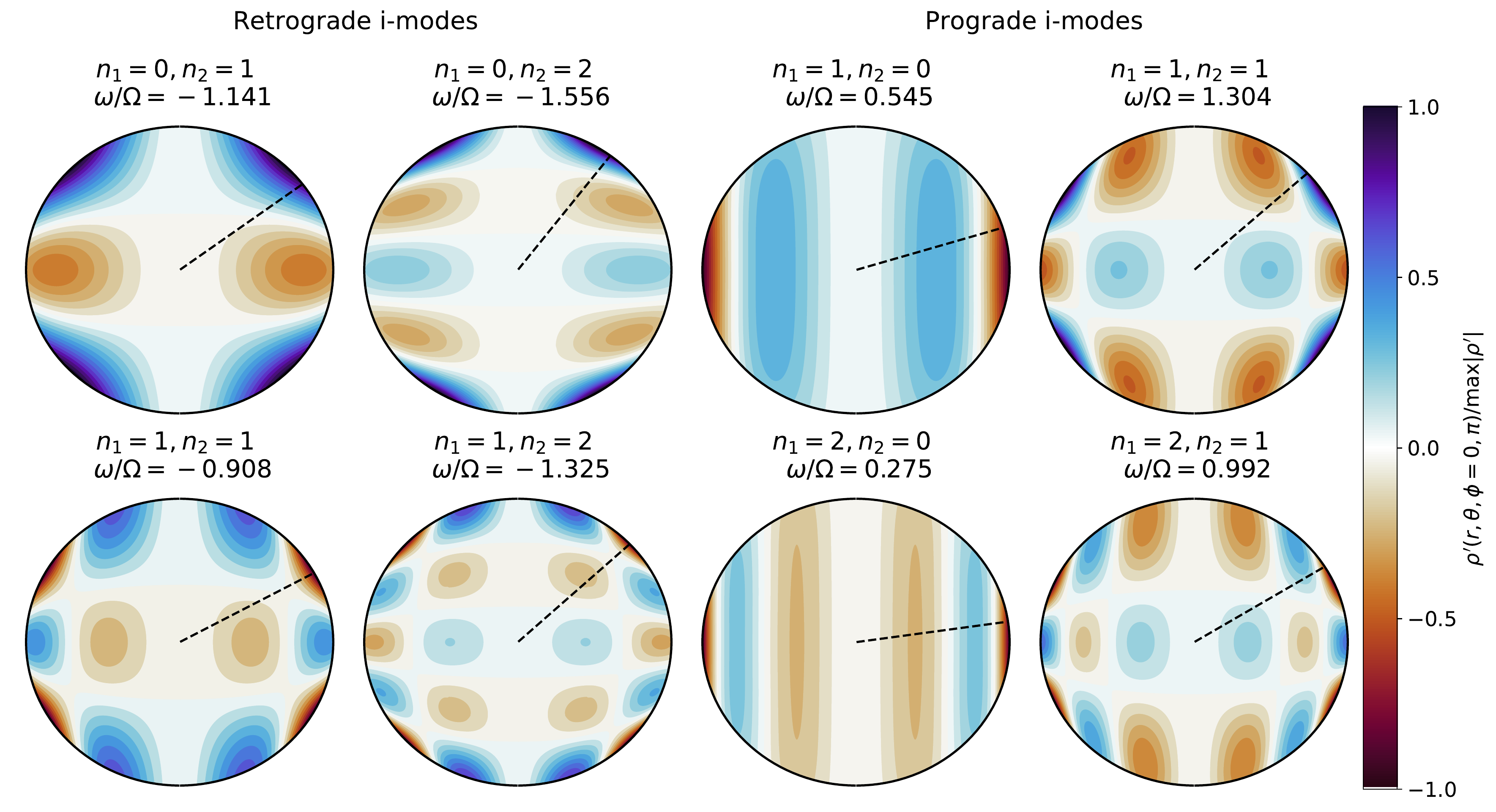}
    \caption{Color-plots showing (arbitrarily normalized) meridional cross-sections of perturbed density $\rho'(r,\theta,\phi=0,\pi)$ for $m=2$ inertial modes with even equatorial parity, calculated from an $n=1$ polytropic model rotating at $\Omega/\Omega_d\simeq0.30$. The mode numbers $n_1$ and $n_2$ describe the number of nodes in spherical generalizations of cylindrical $R$ and $z$ (resp.). The dashed line in each panel indicates the critical polar angle $\theta=\arccos[|\omega|/(2\Omega)]$, which pinpoints the maximum amplitude of the mode's surface eigenfunction in the WKB limit. 
    }
    \label{fig:n1pOm3_imode}
\end{figure*}

\subsection{Inertial modes}\label{sec:imode}
Rotating isentropic stellar models also support global inertial modes (i-modes) that are restored by the Coriolis force, and occupy the low-frequency regime $|\omega|<2\Omega$  \citep{Bryan1889,Greenspan1968}. In the absence of a solid core and/or stable stratification, i-modes form a regular, dense spectrum that can be characterized analytically in the simplest cases: ignoring centrifugal distortion, the i-modes of a rotating fluid body with a homogeneous or power-law density profile have $\rho'\propto P_\ell^m(x_1)P_\ell^m(x_2),$ where $P_\ell^m$ are Legendre polynomials, and $(x_1,x_2)$ are ellipsoidal coordinates that depend on the parameter $\omega/(2\Omega)$ \citep{Wu2005a}. \citet{Lindblom1999} and \citet{Braviner2014} provided similar analytical solutions for the i-modes (and f-modes) of non-spherical Maclaurin spheroids, while \citet{Ivanov2010} and \citet{Papaloizou2010} investigated inertial modes via asymptotic and numerical approaches (respectively).

For each $\ell$ and $m$ in eigensolutions $\propto P_\ell^m(x_1)P_\ell^m(x_2),$ \citet{Wu2005a} found $\ell-m$ eigenfrequencies. The different frequencies correspond to meridional eigenfunctions with differing numbers of nodes with respect to ellipsoidal generalizations of both cylindrical radius $R$ ($n_1$), and cylindrical $z$ ($n_2$). These mode numbers satisfy $\ell-m=2(n_1+n_2)$ for i-modes with even equatorial parity, and $\ell-m=2(n_1+n_2)-1$ for odd equatorial parity. We adopt the labelling scheme $i_{m n_1n_2}^\pm$,\footnote{
Inertial modes can also be labelled by spherical harmonic degrees and azimuthal wavenumbers $m$ \citep[e.g.,][]{Lockitch1999,Xu2017}. We adopt the quasi-cylindrical node numbers $n_1$ and $n_2$ because we find their geometrical interpretation intuitive.
} where $+$ and $-$ again refer to prograde and retrograde propagation. Our focus on modes with even equatorial parity excludes so-called `r-modes,' a subset of odd-parity, retrograde inertial modes \citep[e.g.,][]{Schenk2002}. These r-modes cannot be excited in co-planar systems, but may be relevant when there is a significant spin-orbit misalignment \citep{Ho1999,Braviner2015,Xu2017}.

The meridional cross sections in \autoref{fig:n1pOm3_imode} show $\rho'(r,\theta)$ for even-parity inertial modes with $n_1,n_2\in[0,2]$, calculated for an $n=1,$ $\Omega/\Omega_d\simeq0.3$ polytrope. We see that $n_1$ ($n_2$) can also be counted as the number of surface zeros between the pole (equator) and a `critical latitude' at which the mode wavelengths shortens, and the amplitude approaches a maximum. For each mode, the black dashed line indicates the WKB approximation $\theta=\arccos[|\omega|/(2\Omega)]$ to this critical latitude.

The i-modes with the largest $Q_{\ell m}^\alpha$ are those with the longest wavelengths (i.e., smallest $n_1+n_2$), in particular the oscillations $i_{m01}^-$ and $i_{m10}^+$ with $n_1+n_2=1$ (the first and third cross sections in the top row of \autoref{fig:n1pOm3_imode}). For convenience, we sometimes refer to these oscillations as \emph{the} retrograde and prograde inertial modes, respectively \citep[for comparison, $i_{201}^-$ and $i_{210}^+$ are the same as the $j=3$, $m=2$ modes described in table IV of][]{Xu2017}. For completeness, for each $m$ we additionally consider the four shorter-wavelength modes with $n_1+n_2=2$.

We note that the presence of a rigid core complicates inertial wave propagation \citep{Rieutord1997,Ogilvie2005,Goodman2009}, by introducing singularities on the inner boundary that disrupt the regular spectrum illustrated by \autoref{fig:n1pOm3_imode}. In the (apparent) absence of the global inertial modes of coreless models, inertial wave attractors do appear as spatially periodic features. However, these attractors do not generally correspond to the largest peaks in (frequency-dependent) tidal dissipation \citep{Ogilvie2009,Rieutord2010,Papaloizou2010,Ogilvie2013}. 

Instead, \citet{Lin2021} have recently shown that the largest peaks in dissipation in rotating shells can still be associated with resonances involving large-scale, smooth flows. These flows are reminiscent of the global inertial modes of full spheres and spheroids, but are hidden beneath wave beams launched at critical latitudes on the core. The findings of \citet{Lin2021} provide further motivation for considering the pure inertial modes of coreless, centrifugally distorted models, even if a given giant planet may have a solid core (although tidal overlap coefficients may differ between planetary models that do and do not include a core).

\section{Results}\label{sec:fires}
In this section we present the results of our mode and Love number calculations for rapidly rotating, $n=1$ polytropes. Section \ref{sec:mrel} demonstrates the relative importance of the different modes described in Section \ref{sec:modes} to the tidal response, Section \ref{sec:frQeps} characterizes the variation of mode frequencies and tidal coefficients with increasing rotation rate, and Section \ref{sec:klmcalc} presents representative Love number calculations relevant to tides in rotating gas giants. Lastly, Section \ref{sec:rapid} describes additional results for more rapid rotation.

\begin{figure*}
    \centering
    \includegraphics[width=\columnwidth]{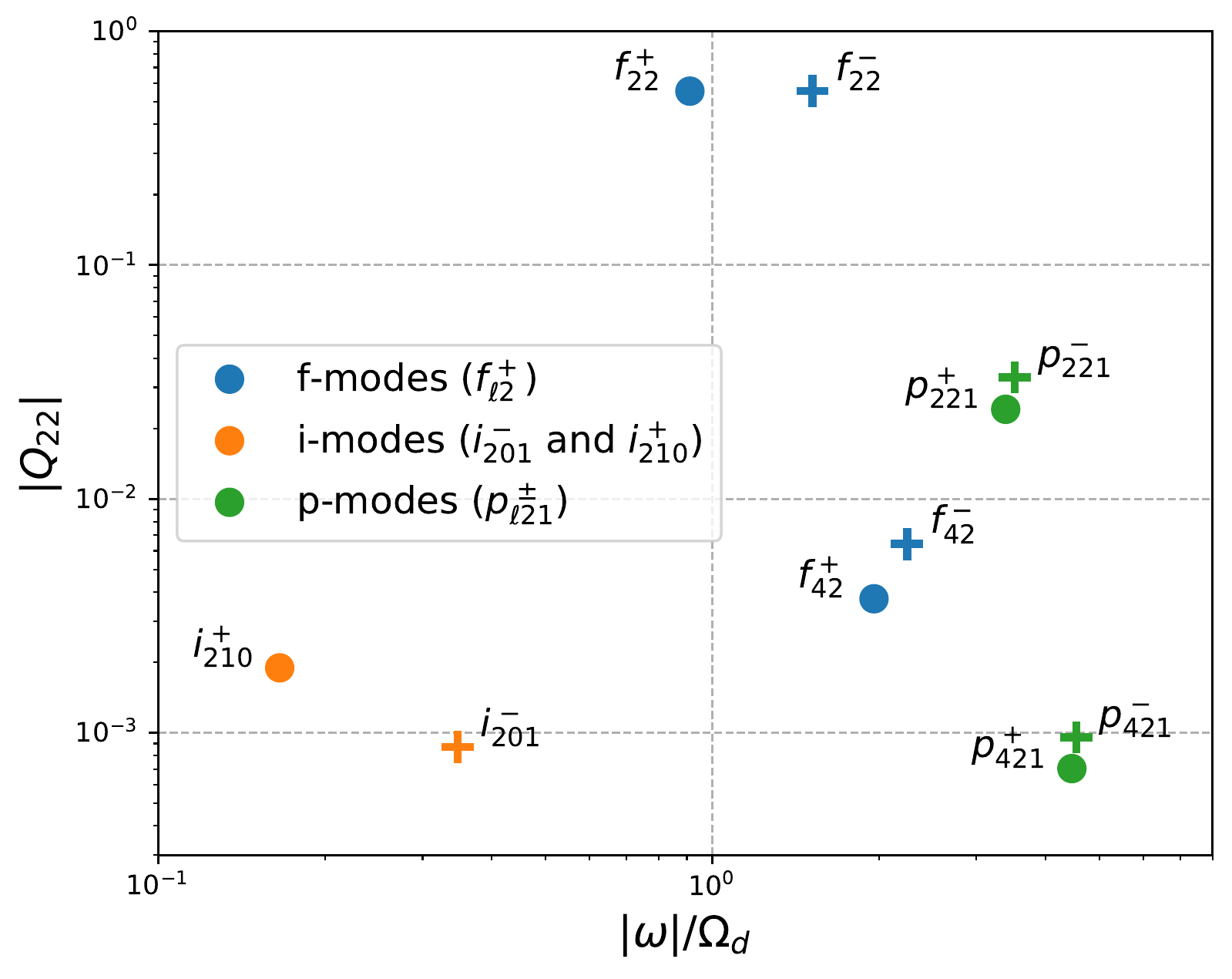}
    \includegraphics[width=\columnwidth]{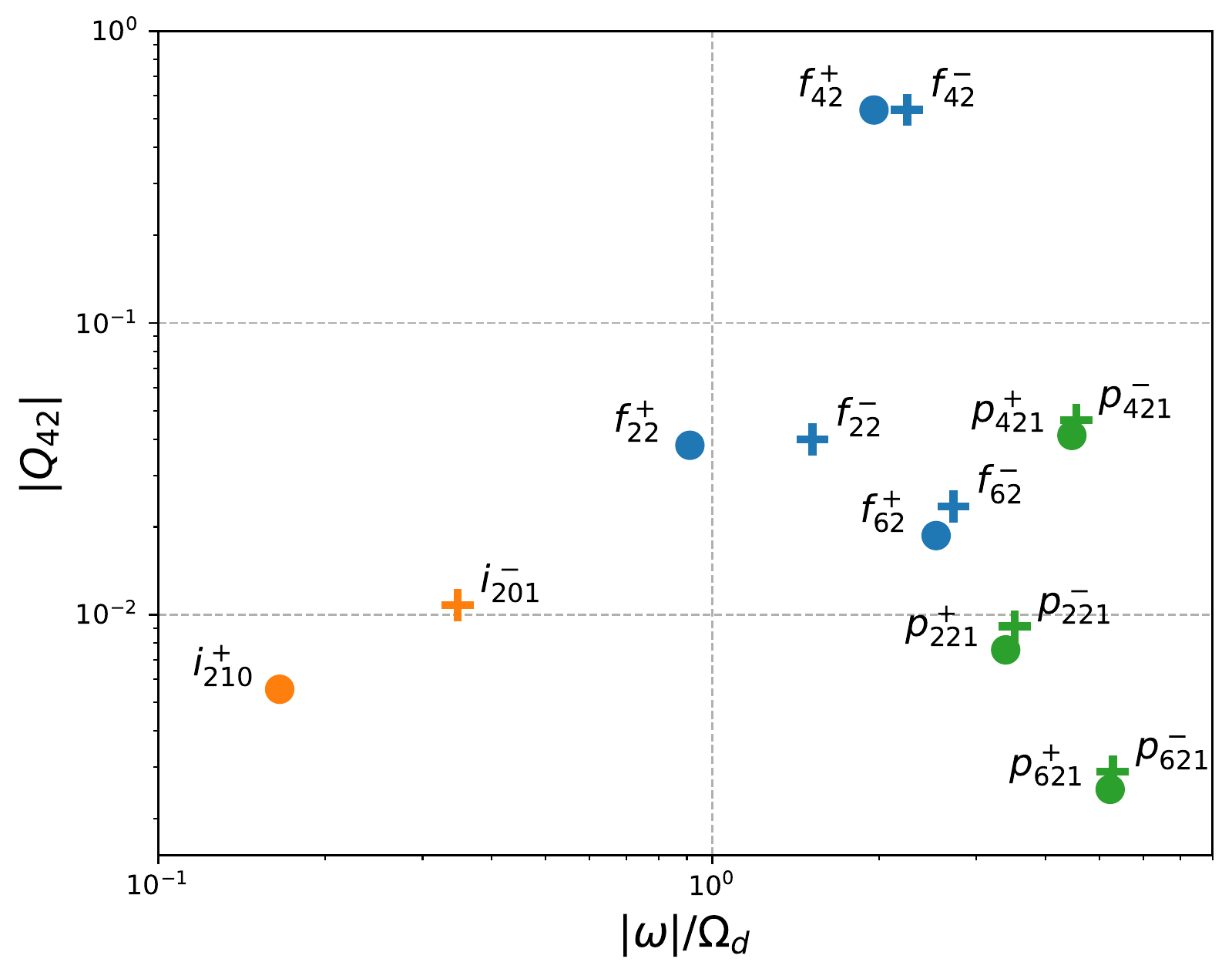}
    \caption{Plots showing $|Q_{22}|$ (left) and $|Q_{42}|$ (right) vs. frequency magnitude $|\omega|$ for a selection of $m=2$ f-modes, p-modes and i-modes calculated for an $n=1$ polytropic model with rotation rate $\Omega/\Omega_d\simeq0.3$. Filled circles and plus signs denote prograde and retrograde modes, respectively. We label each point according to the conventions described in Section \ref{sec:modes}: $f_{\ell m}^\pm$ for f-modes, $p_{\ell mn_p}^\pm$ for p-modes, and $i_{mn_1n_2}^\pm$ for i-modes.}
    \label{fig:n1pOm3Q}
\end{figure*}

\subsection{Mode relevance}\label{sec:mrel}
The first question to ask is which modes contribute most significantly to which Love numbers. As might be gleaned from \autoref{eq:klma}, mode contributions $k_{\ell m}^\alpha$ to a given $k_{\ell m}$ are largest for i) large $Q_{\ell m}^\alpha,$ ii) small $\epsilon_\alpha$, and iii) frequencies $\omega_\alpha$ close to resonance with $\omega_m$ (i.e., small $\omega_\alpha-\omega_m$). Since $\epsilon_\alpha$ is approximately equal to the zero-rotation mode frequency (for f and p-modes), i) and iii) prove to be the differentiating factors. \autoref{fig:n1pOm3Q} demonstrates the relative importance of different oscillations to an $n=1$ polytrope rotating wwith $\Omega/\Omega_d\simeq0.3$. The left and right-hand panels plot $|Q_{22}^\alpha|$ and $|Q_{42}^\alpha|$ (resp.) vs. $|\omega_\alpha|$ for a selection of f-modes (blue), p-modes (green) and i-modes (orange). The filled circles (plus signs) indicate prograde (retrograde) modes with positive (negative) $\omega_\alpha$. 

The plots show that for a rotation rate typical of gas giants ($\Omega/\Omega_d\simeq 0.3,0.4$ respectively for Jupiter and Saturn) the f-modes $f_{\ell m}^\pm$ (associated with spherical harmonic $Y_\ell^m$ in the absence of rotation) produce $Q_{\ell m}^\alpha$ orders of magnitude larger than any of the other modes. P-modes possess reasonably large $Q_{\ell m}^\alpha$, but have large frequencies that are unlikely to come anywhere near resonance with satellites. Inertial modes, on the other hand, have smaller $Q_{\ell m}^\alpha$ that can nevertheless be compensated by frequencies falling much closer to resonance with realistic tidal forcing frequencies.

\autoref{fig:n1pOm3Q} also demonstrates coupling across spherical harmonics by rotation, which produces non-zero $Q_{\ell m}^\alpha$ for degrees other than that associated with each mode's non-rotating $\ell$. For instance, when $\Omega=0$ the only non-zero tidal coupling factor for $f_{22}^{\pm}$ is $Q_{22},$ but with rotation the modes have non-zero $Q_{42},Q_{62},...$ that decrease in magnitude with increasing $\ell$ (see \autoref{app:QvMode} for a more detailed breakdown of $Q_{\ell m}^\alpha$ calculations for different modes). As the f-mode eigenfunctions come to involve more spherical harmonics, they overlap more strongly with multiple components of the tidal potential. For moderate to rapid rotation rates, a given $k_{\ell m}$ can no longer be reasonably approximated through sole consideration of the corresponding $f_{\ell m}^\pm,$ since every mode overlaps non-negligibly with multiple components of both the tidal potential and the induced response. 

Comparison of the left and right panels in \autoref{fig:n1pOm3Q} reveals an important aspect of inertial mode overlap with the tidal potential: i-modes of a given $m$ couple most strongly to tesseral harmonics with degrees $\ell>m$. For instance, the longest wavelength $m=2$ i-modes ($i_{201}^-$ and $i_{210}^+$) produce larger $|Q_{42}|$ than $|Q_{22}|.$ This preferential coupling with tesseral harmonics is exact for homogeneous, slowly rotating fluid bodies \citep{Ogilvie2013}. The i-modes of the distorted $n=1$ polytropes considered here have nonzero sectoral overlap, but the preferential coupling with tesseral harmonics still holds approximately, leading to tesseral overlap coefficients that are larger by at least an order of magnitude.

\begin{figure*}
    \centering
    \includegraphics[width=.66\columnwidth]{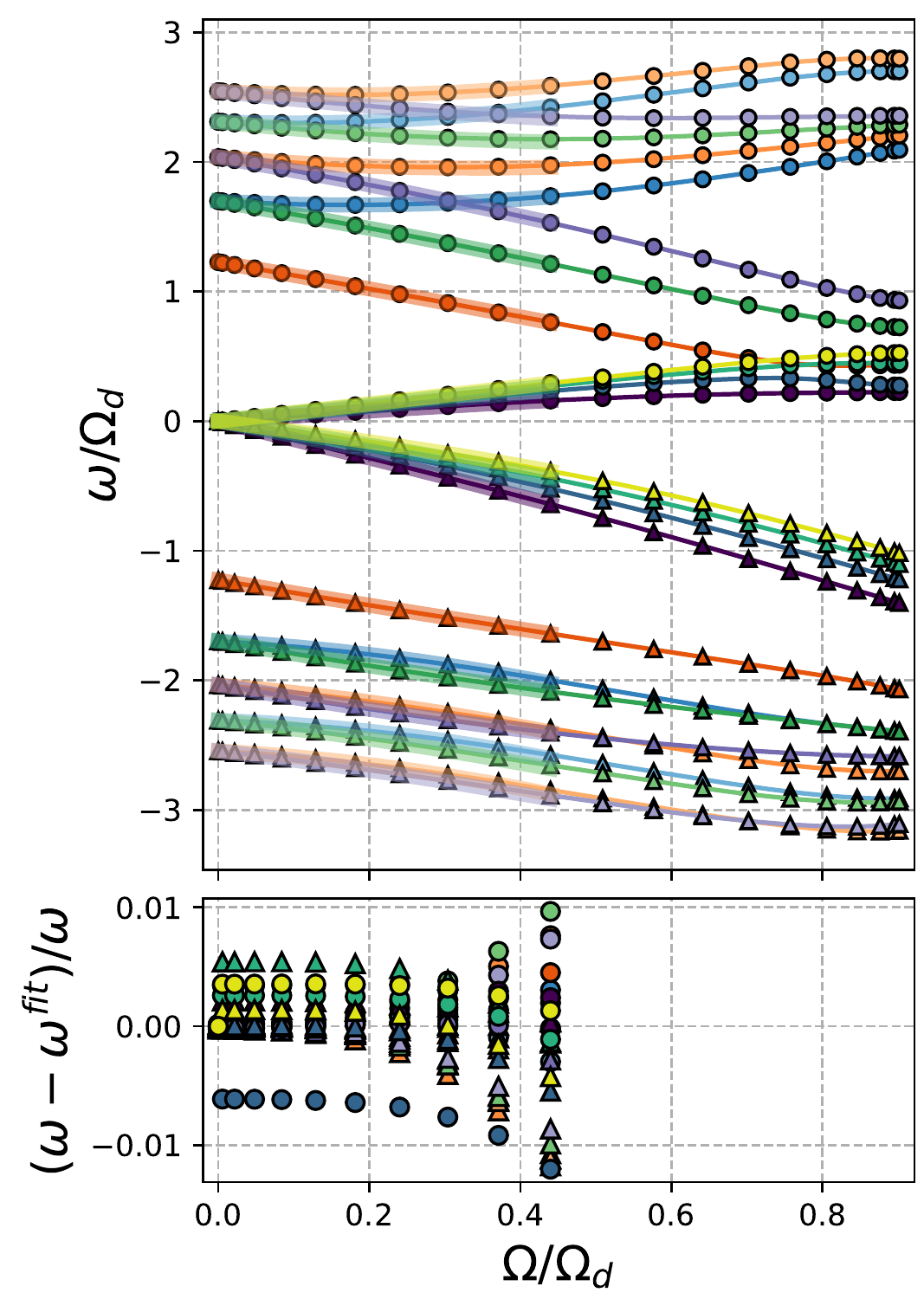}
    \includegraphics[width=.66\columnwidth]{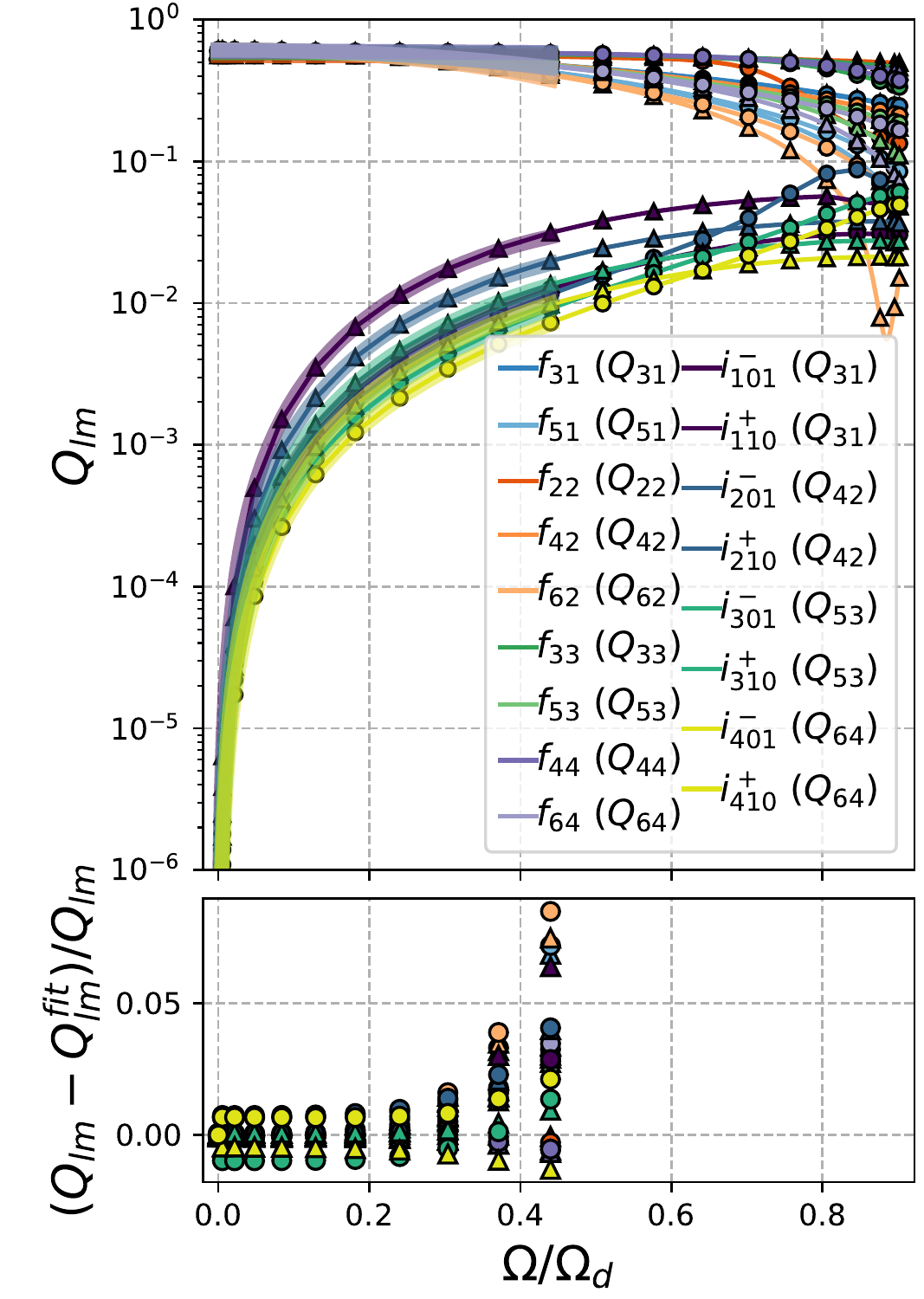}
    \includegraphics[width=.66\columnwidth]{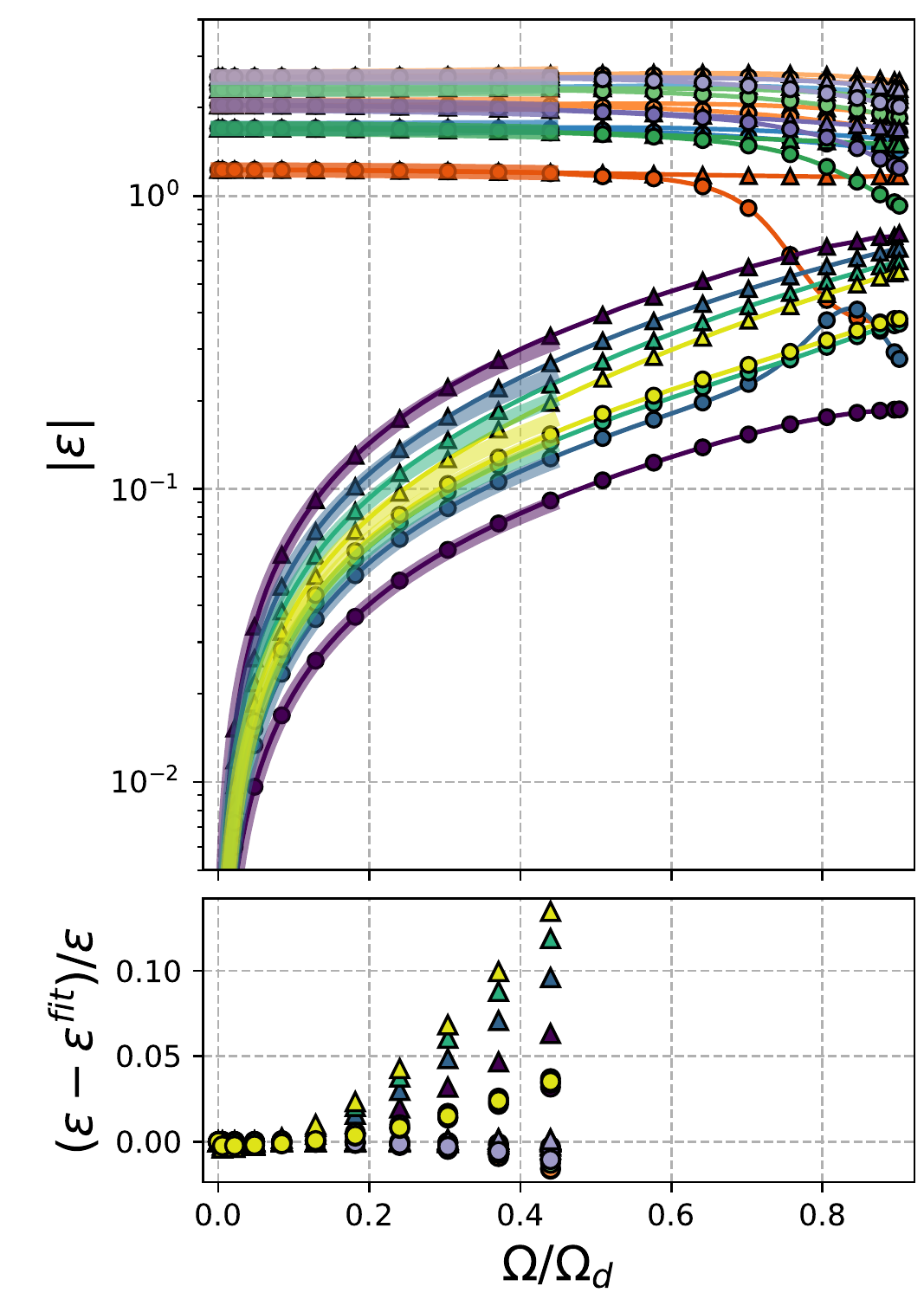}
    \caption{Top panels: oscillation frequencies (left), dominant tidal overlap coefficients $Q_{\ell m}^\alpha$ (center), and $\epsilon$ coefficients calculated for a subset of f-modes and i-modes of $n=1$ polytropes for different rotation rates. The filled circles and triangles denote prograde and retrograde modes, respectively. The thick lines illustrate least-squares polynomial fits out to $\Omega/\Omega_d\simeq0.43$ (provided in Table \ref{tab:n1fmode} and Table \ref{tab:n1imode}). Bottom panels: residuals associated with the polynomial fits. Due to the dominance of inertial mode coupling to tesseral (rather than sectoral) harmonics, in the center panels we plot $Q_{m+2,m}$ (rather than $Q_{mm}$) for the i-modes.}
    \label{fig:n1poly_omQlm} 
\end{figure*}

\subsection{Frequency and overlap variation with rotation}\label{sec:frQeps}
The top panels in \autoref{fig:n1poly_omQlm} plot frequencies $\omega$ (left), dominant tidal overlap coefficients $Q_{\ell m}$ (middle), and $\epsilon$ coefficients (right) as a function of rotation rate for several f-modes and the longest wavelength ($n_1+n_2=1$) i-modes of the $n=1$ polytrope. In this section we present analytic formulae (polynomial fits) to these quantities as a function of rotation rate.

\subsubsection{Fits for f-modes}
The frequencies and $\epsilon$ parameters of f-modes ($f_{\ell m}^\pm$) can be expanded to order $\Omega^2$ as 
\begin{align}\label{eq:fmfit1}
    \omega&\simeq\omega_0+A\Omega+B\Omega^2,
\\\label{eq:fmfit2}
    \epsilon&\simeq\omega_0+D\Omega^2,
\end{align}
where $\omega_0$ is the non-rotating frequency. Note that $\epsilon$ has no linear dependence on $\Omega$. The linear term $A\Omega$ arises due to the Coriolis force, and can be obtained by treating rotation as a perturbation. Defining the Lagrangian displacement for a mode $\alpha$ in the absence of rotation as 
$\boldsymbol{\xi}_\alpha^0=(\xi_r\hat{\bf r} + r\xi_\perp\nabla_\perp)Y_\ell^m$ 
(here $\nabla_\perp$ is the angular part of the gradient operator), with the normalization $\langle \boldsymbol{\xi}_\alpha^0,\boldsymbol{\xi}_\alpha^0\rangle=1$, the perturbative approach yields \citep[e.g.,][]{Unno1989}
\begin{equation}
    A=-m\int_0^R r^2\rho_0(2\xi_r\xi_\perp + \xi_\perp^2)\text{d}r.
\end{equation}

The coefficient describing the overlap between $f_{\ell m}^\pm$ and the $\ell m$-tidal potential can be expanded as
\begin{align}\label{eq:fmfit3}
    Q_{\ell m}&\simeq Q_0+C\Omega^2,
\end{align}
where $Q_0$ is the tidal overlap in the absence of rotation. Table \ref{tab:n1fmode} gives values obtained for $A,B,C,D$ by fitting our numerical calculations for $\Omega/\Omega_d\lesssim0.43$. The thick lines in \autoref{fig:n1poly_omQlm} plot these polynomial fits, while the bottom panels show residuals between the fits and our numerical calculations. The numerical values in Table \ref{tab:n1fmode} are qualitatively consistent with the results of \citet{Barker2016}, who found that for homogeneous spheroids the relative importance of centrifugal distortion to changes in f-mode frequencies (i.e., the ratio $B/A$) increases with $\ell$ \citep[see also Section 3.1 of ][]{Ho1999}.
\begin{deluxetable}{cccccccccc}
\tablecaption{
    Fits to f-mode frequencies $\omega\simeq\omega_0 + A\Omega + B\Omega^2$, tidal overlap coefficients $Q_{\ell m}^\alpha\simeq Q_0 + C\Omega^2$, and coefficients $\epsilon\simeq\omega_0+D\Omega^2$. The mode notation is $f_{\ell m}^\pm$ (see Section \ref{sec:fmode}). All quantities are in units with $G=M=R_\text{eq}=1$.
}\label{tab:n1fmode}
\tablewidth{0pt}
\tablehead{
    \colhead{Mode} & 
    \colhead{$|\omega_0|$} &  
    \colhead{A} &
    \colhead{B} &
    \colhead{$Q_0$} &
    \colhead{C} &
    \colhead{D} 
}
\startdata
    $f_{31}^\pm$ & $\pm1.698$ & $-0.33$ & $\pm0.92$ & 
        $0.585$ & $-0.62$ & $\pm0.08$ \\
    $f_{51}^\pm$ & $\pm2.310$ & $-0.20$ & $\pm1.07$ & 
        $0.606$ & $-1.08$ & $\pm0.33$ \\
    \hline
    $f_{22}^\pm$ & $\pm1.227$ & $-1.00$ & $\mp0.14$ & 
        $0.558$ & $-0.04$ & $\mp0.14$ \\
    $f_{42}^\pm$ & $\pm2.037$ & $-0.50$ & $\pm0.74$ & 
        $0.598$ & $-0.70$ & $\pm0.11$ \\
    $f_{62}^\pm$ & $\pm2.546$ & $-0.33$ & $\pm0.97$ & 
        $0.610$ & $-1.21$ & $\pm0.33$ \\
    \hline
    $f_{33}^\pm$ & $\pm1.698$ & $-1.00$ & $\mp0.24$ & 
        $0.585$ & $-0.06$ & $\mp0.25$ \\
    $f_{53}^\pm$ & $\pm2.310$  & $-0.60$ & $\pm0.56$ & 
        $0.606$ & $-0.75$ & $\pm0.07$ \\
    \hline
    $f_{44}^\pm$ & $\pm2.037$ & $-1.00$ & $\mp0.34$ & 
       $0.598$ & $-0.08$ & $\mp0.35$ \\
    $f_{64}^\pm$ & $\pm2.546$ & $-0.66$ & $\pm0.41$ &
        $0.610$ & $-0.80$ & $\pm0.00$ \\
\enddata
\end{deluxetable}

As noted in Section \ref{sec:mrel}, for $\Omega>0$ a given f-mode $f_{\ell m}^\pm$ also couples to tidal potentials associated with different spherical harmonic degrees ($\ell\pm2,\ell\pm4,..$). In the non-rotating limit the sign of $Q_0$ is unimportant, since only its square appears in \autoref{eq:klma}. For a rapidly rotating planet, however, the (relative) signs of the different $Q_{\ell m}^\alpha$ associated with a mode $\alpha$ must be retained, since the sum over degrees in \autoref{eq:klma} includes cross terms such as $Q_{\ell m}^\alpha Q_{\ell\pm2,m}^\alpha$. Adopting the convention that $Q_0>0$, we find that the most important of the subdominant tidal overlap coefficients are the tesseral $Q_{m+2,m}$ associated with the sectoral f-modes $f_{mm}^\pm.$ Fitting our numerical results, we find
\begin{align}
    Q_{42}(f_{22}^\pm)
    &\simeq -0.46\Omega^2, \\
    Q_{53}(f_{33}^\pm)
    &\simeq -0.56\Omega^2, \\
    Q_{64}(f_{44}^\pm)
    &\simeq -0.63\Omega^2.
\end{align}
The negative values of these overlap coefficients (relative to $Q_0$) are important to include, along with the signs of the $W_{\ell m}$ coefficients. The additional overlap coefficients for non-sectoral f-modes also scale with $\Omega^2$ [e.g., $Q_{22}(f_{42}^\pm)\simeq E\Omega^2$, where $E\ll1$], but we find that their contributions to the relevant Love numbers are negligible.

\begin{deluxetable*}{cccccccc}
\tablecaption{
    Fits to i-mode frequencies $\omega\simeq A\Omega + B\Omega^3$, tidal overlap coefficients $Q_{mm}\simeq  C\Omega^2+D\Omega^4$ and $Q_{m+2,m}\simeq  E\Omega^2+F\Omega^4$, and coefficients $\epsilon\simeq G\Omega$. The mode notation is $i_{mn_1n_2}^\pm$ (see Section \ref{sec:imode}). We do not list $Q_{mm}$ for the two $m=1$ i-modes because the tidal potential does not include an $\ell=m=1$ component.
}\label{tab:n1imode}
\tablewidth{0pt}
\tablehead{
    \colhead{Mode} & 
    \colhead{A} &  
    \colhead{B} &
    \colhead{C} &
    \colhead{D} &
    \colhead{E} &
    \colhead{F} &
    \colhead{G} 
}
\startdata
    $i_{110}^+$ & $+0.40$ & $-0.20$ & -- & -- & $+0.074$ & $-0.056$ & $+0.201$  \\
    $i_{101}^-$ & $-1.41$ & $-0.24$ & -- & -- & $+0.217$ & $-0.338$ & $-0.708$ \\
    \hline
    $i_{210}^+$ & $+0.56$ & $-0.12$ & $-0.015$ & $-0.049$ & $+0.061$ & $-0.019$ & $+0.279$ \\
    $i_{201}^-$ & $-1.10$ & $-0.46$ & $+0.010$ & $-0.004$ & $+0.130$ & $-0.169$ &  $-0.552$ \\
    \hline
    $i_{310}^+$ & $+0.63$ & $-0.06$ & $-0.013$ & $-0.032$ & $+0.048$ & $-0.007$ & $+0.316$ \\
    $i_{301}^-$ & $-0.90$ & $-0.54$ & $+0.006$ & $+0.000$ & $+0.085$ & $-0.081$ & $-0.454$ \\
    \hline
    $i_{410}^+$ & $+0.67$ & $-0.02$ & $-0.010$ & $-0.024$ & $+0.037$ & $-0.001$ & $+0.337$ \\
    $i_{401}^-$ & $-0.77$ & $-0.56$ & $+0.004$ & $+0.002$ & $+0.059$ & $-0.040$ & $-0.387$ \\
\enddata
\end{deluxetable*}

\subsubsection{Fits for i-modes}
Table \ref{tab:n1imode} provides coefficients for polynomial fits to the frequencies, overlap coefficients $Q_{mm}$ and $Q_{m+2,m}$, and the $\epsilon$-coefficients of the longest-wavelength inertial modes with even equatorial parity ($i_{m01}^-$ and $i_{m10}^+$). We find that these quantities are represented accurately out to $\Omega/\Omega_d\simeq0.43$ by fits with the form
\begin{align}
    \omega&\simeq A\Omega+B\Omega^3,\\
    Q_{mm}&\simeq C\Omega^2+D\Omega^4,\\
    Q_{m+2,m}&\simeq E\Omega^2+F\Omega^4,\\
    \epsilon&\simeq G\Omega.
\end{align}
Note that the coefficients in these expansions are distinct from those introduced in Equations \eqref{eq:fmfit1}, \eqref{eq:fmfit2}, and \eqref{eq:fmfit3}. We choose the (larger amplitude) $Q_{m+2,m}$ to be positive. As indicated by the table, this leads to negative $Q_{mm}$ for the prograde inertial modes, and positive $Q_{mm}$ for the retrograde inertial modes.

Our results for the $m=2$ i-modes agree with those given in \citet{Xu2017} (compare with the $j=3$, $m=2$ modes in their Table IV), although they did not consider the (somewhat larger) $Q_{42}$ coefficients. To further compare with \citet{Xu2017}, we have computed a few $m=1$ inertial modes with odd equatorial parity, namely the $m=1$ r-mode (labelled $i_{1r}^-$), and the three $m=1$, odd-parity i-modes corresponding to their $j=3,$ $m=1$ modes. Note that because these modes have odd equatorial parity, they cannot couple to the tidal potential for systems with spin-orbit alignment (the relevant case for Jupiter and Saturn). 

Our results, given in Table \ref{tab:iom}, show agreement with those of \citet{Xu2017}. Note that the $m=1$ r-mode has a frequency exactly equal to $-\Omega$ (to the precision of our calculations) for all rotation rates considered in this paper. As explained by \citet{Xu2017}, this implies that the mode has zero inertial-frame frequency, constituting a spin-over perturbation.

\begin{deluxetable}{cccccccc}\label{tab:iom}
\tablecaption{
    Same as Table \ref{tab:iom}, but for odd-parity, $m=1$ i-mode frequencies $\omega\simeq A\Omega + B\Omega^3$, tidal coupling factors $Q_{21}\simeq C\Omega^2+D\Omega^4$, and coefficients $\epsilon\simeq E\Omega$.
}
\tablewidth{0pt}
\tablehead{
    \colhead{Mode} & 
    \colhead{A} &  
    \colhead{B} &
    \colhead{C} &
    \colhead{D} &
    \colhead{E} 
}
\startdata
    $i_{1r}^-$ & $-1.000$ & $0.000$  & $+0.370$ & $-0.276$ & $-0.500$ \\
    $i_{102}^-$ & $-1.613$ & $-0.269$ & $+0.014$ & $-0.027$ & $-0.807$  \\
    $i_{111}^-$ & $-0.690$ & $-0.361$ & $+0.009$ & $+0.011$ & $-0.345$  \\
    $i_{111}^+$ & $+1.032$ & $+0.211$ & $+0.017$ & $+0.036$ & $+0.516$ \\
\enddata
\end{deluxetable}

\subsection{Love number calculations}\label{sec:klmcalc}
\begin{figure}
    \centering
    \includegraphics[width=\columnwidth]{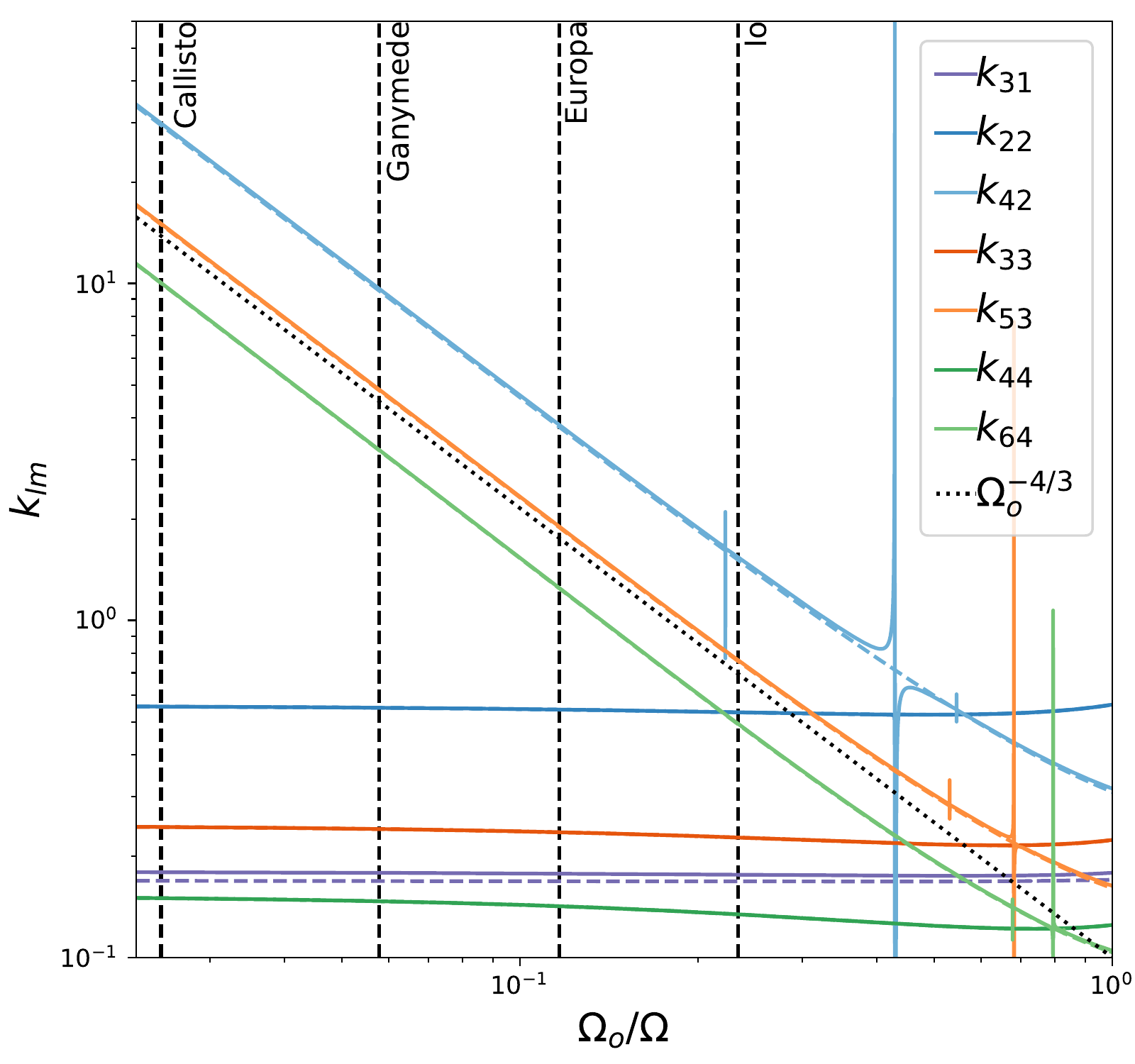}
    \caption{Love numbers $k_{\ell m}$ as a function of orbital frequency $\Omega_o$ (in units of the planet's rotation frequency $\Omega$), computed via \autoref{eq:klma} 
    for an $n=1$ polytrope with $\Omega/\Omega_d\simeq0.30$. The dashed colored lines show Love numbers computed including only  the contributions of f-modes and p-modes, while the solid colored lines show results including f-modes, p-modes, and i-modes. The vertical black dashed lines indicate tidal forcing frequencies associated with the Galilean satellites. The dotted black line illustrates a power law $\propto\Omega_o^{-4/3}$ that derives from \autoref{eq:km2m}, and is associated with a dominant contribution of sectoral f-modes driven by the sectoral tide to the tesseral tidal response. Resonances between i-modes and the tidal forcing frequency produce formally infinite peaks that would gain a finite height with the inclusion of dissipation and/or nonlinearity.
    }
    \label{fig:n1poly_klm_modevar} 
\end{figure}

\autoref{fig:n1poly_klm_modevar} plots a variety of tidal Love numbers $k_{\ell m}$, computed via \autoref{eq:klma} using our calculations of mode frequencies and tidal overlap coefficients for 
an $n=1$ polytrope with a (roughly Jupiter-appropriate) rotation rate of $\Omega/\Omega_d\simeq0.30$, as a function of $\Omega_o/\Omega$. The dashed lines show results computed including contributions from only f-modes and p-modes, while the solid lines show results computed with all modes (f,p and i-modes). These curves can be closely reproduced using \autoref{eq:klma} and the fits provided in Tables \ref{tab:n1fmode} and \ref{tab:n1imode} (excluding p-modes minimally affects our results). 

The so-called `hydrostatic' Love numbers can be computed by taking the limit of zero tidal forcing frequency $\omega_m=m(\Omega_o-\Omega)=0$, i.e., 
\begin{equation}
    k_{\ell m}^\text{hs}=k_{\ell m}(\Omega_0/\Omega=1).
\end{equation}
This quantity is \emph{satellite independent} in the limit of negligible mass ratio $q=M'/M$, characterizing the intrinsic (linear) response of the planet to a hypothetical perturber with infinitesimal mass and an orbit synchronized to the spin rate. 

This definition differs from that used in studies employing the CMS method that report hydrostatic Love numbers for multiple satellites; we reserve the term hydrostatic for the response that is stationary in the co-rotating frame, which for circular and co-planar systems with vanishing $q$ can only be produced by a single orbital frequency $\Omega_o$ and associated satellite separation $a$. A more direct comparison of our results to those of, e.g., \citet{Wahl2017,Wahl2020} could be made by assuming $\omega_m=0$ in \autoref{eq:klma}, but then introducing multiple values of $a$ in \autoref{eq:Uratio} (this is inconsistent within our framework because fixing $\omega_m=0$, and hence $\Omega_o=\Omega$, in turn fixes $a$).

The analytically known $k_{22}^\text{hs}=15/\pi^2-1$ of a non-rotating, $n=1$ polytrope provides a zeroth order test of our computations; including only contributions from $f_{22}^\pm$ and $p_{221}^\pm,$ we find an error of $\lesssim0.0001\%$. Table \ref{tab:klm} lists hydrostatic Love numbers taken from the calculations shown in \autoref{fig:n1poly_klm_modevar} for a variety of $\ell$ and $m,$ in addition to frequency-dependent values obtained by identifying intersections with the forcing frequencies associated with the Galilean satellites.

Of the hydrostatic values shown in Table \ref{tab:klm}, the tesseral Love numbers with $\ell>m$ and $m>1$ (e.g., $k_{42}^\text{hs}$) exhibit the largest deviations from the hydrostatic values found by \citet{Lai2021} with a perturbative $\sim{\cal O}(\Omega)$ treatment of rotation. However, the hydrostatic values of $k_{\ell m}^\text{hs}$ computed both in this work and by \citet{Lai2021} are dwarfed by the tesseral Love numbers' variation with frequency. To elucidate this frequency dependence, in \autoref{fig:n1poly_dklm_modevar} we plot dynamical Love number corrections given by
\begin{equation}
    \frac{\Delta k_{\ell m}}{k_{\ell m}^\text{hs}}
    =\frac{(k_{\ell m}-k_{\ell m}^\text{hs})}
    {k_{\ell m}^\text{hs}}.
\end{equation}
The top panel shows dynamical corrections for $k_{31}$ and the sectoral Love numbers $k_{mm}$, while the bottom panel shows corrections for the tesseral $k_{m+2,m}$. In the following subsections we break down the contributions of different modes to different Love numbers.

\begin{deluxetable}{cccccccccc}\label{tab:klm}
\tablecaption{
    Love numbers $k_{\ell m}$ computed for an $n=1$ polytrope with $\Omega/\Omega_d\simeq0.30$ ($\sim$ Jupiter's rotation rate), both in the hydrostatic limit of zero forcing frequency (i.e., $\Omega_o=\Omega$), and at the orbital frequencies associated with the Galilean satellites.
}
\tablewidth{0pt}
\tablehead{
    \colhead{} & 
    \colhead{$k_{31}$} &
    \colhead{$k_{22}$} &  
    \colhead{$k_{42}$} &
    \colhead{$k_{33}$} &
    \colhead{$k_{53}$} &
    \colhead{$k_{44}$} &
    \colhead{$k_{64}$}
}
\startdata
    Hydrostatic & 0.18 & 0.56 & 0.32  & 0.22 & 0.16 & 0.12 & 0.10 \\
    Io          & 0.18 & 0.53 & 1.54  & 0.23 & 0.76  & 0.13 & 0.49 \\
    Europa      & 0.18 & 0.54 & 3.81  & 0.24 & 1.90  & 0.14 & 1.25 \\
    Ganymede    & 0.18 & 0.55 & 9.65  & 0.24 & 4.84  & 0.15 & 3.21 \\
    Callisto    & 0.18 & 0.56 & 29.89 & 0.24 & 15.06 & 0.15 & 10.06 \\
\enddata
\end{deluxetable}

\subsubsection{The effect of f-modes on Love numbers}\label{sec:fmneg}
\begin{figure}
    \centering
    \includegraphics[width=\columnwidth]{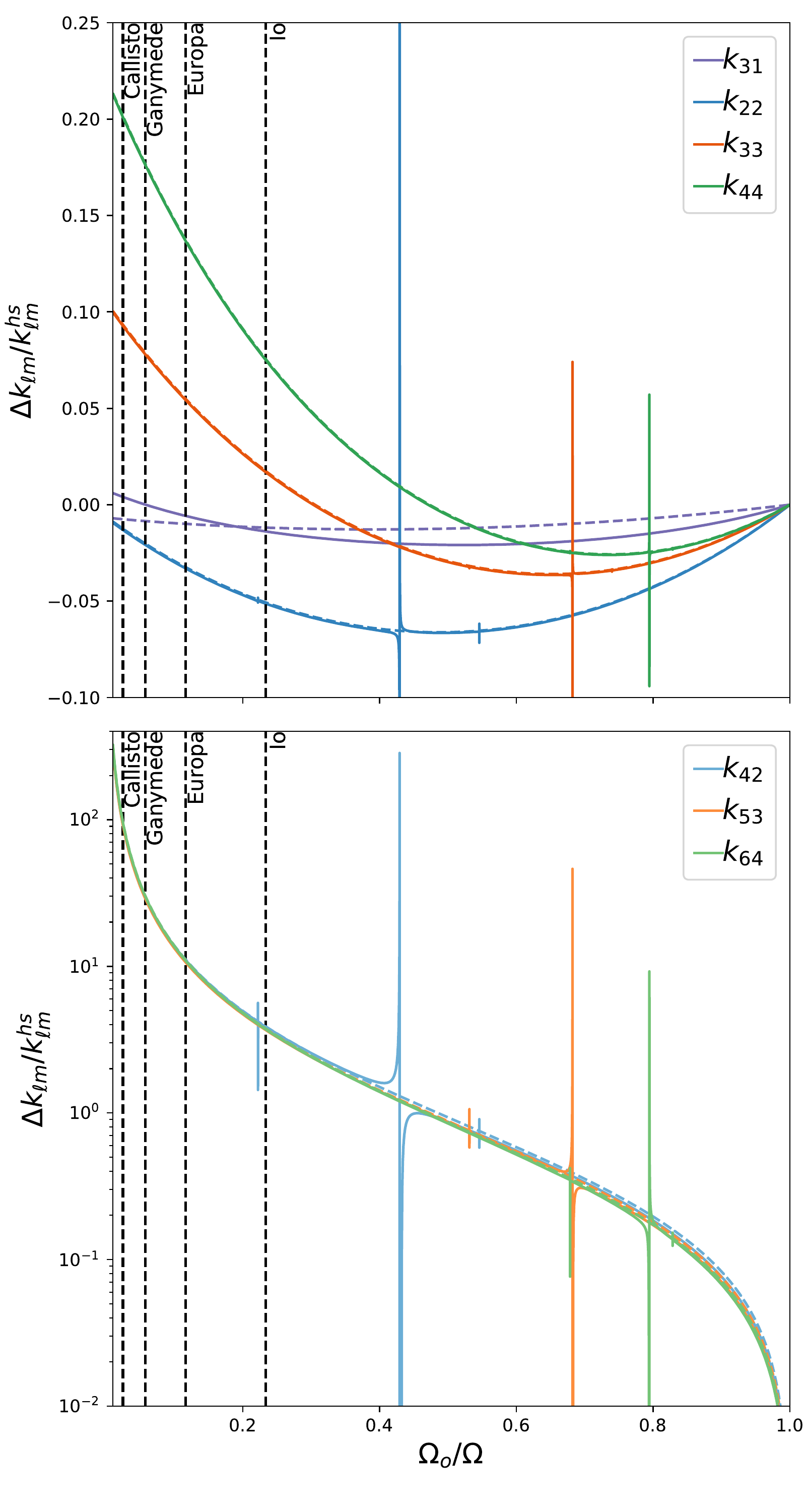}
    \caption{Same as \autoref{fig:n1poly_klm_modevar}, but showing dynamical corrections 
    $\Delta k_{\ell m}/k_{\ell m}^\text{hs}
    =(k_{\ell m}-k_{\ell m}^\text{hs})/k_{\ell m}^\text{hs}$ to $k_{31}$ and the sectoral Love numbers $k_{mm}$ (top), and the tesseral Love numbers $k_{m+2,m}$ (bottom). Inertial modes produce resonant spikes where their frequencies match the tidal forcing frequency (see, e.g., the resonance produced by $i_{201}^-$ near $\Omega_o/\Omega=0.45$). These resonances are wider for the tesseral Love numbers because i-modes couple more strongly to tesseral harmonics.
    }
    \label{fig:n1poly_dklm_modevar} 
\end{figure}

As indicated by the dashed colored lines in \autoref{fig:n1poly_klm_modevar} and \autoref{fig:n1poly_dklm_modevar}, the f-modes contribute relatively gradual frequency variation to $k_{31}$, $k_{22},$ $k_{33}$ and $k_{44}$. The top panel of \autoref{fig:n1poly_dklm_modevar} shows dynamical corrections that follow the same general trends found by \citet{Idini2021} and \citet{Lai2021}; the inclusion of f-mode modification by centrifugal distortion in our computations affects the overall values of these Love numbers (i.e., $k_{\ell m}^\text{hs}$) more significantly than their dynamical corrections. 

In contrast, the f-modes contribute dramatic frequency variation to the tesseral Love numbers with $\ell>m$ (and $m>1$) that cannot be captured by ${\cal O}(\Omega)$ treatments of rotation that neglect centrifugal distortion. The curves for $k_{42},$ $k_{53}$ and $k_{64}$ in the log-log plot shown in \autoref{fig:n1poly_klm_modevar} indicate a power-law dependence $\propto\Omega_o^{-4/3}$ (black dotted line) as $\Omega_o\rightarrow0$. The origin of this power law can be traced back to \autoref{eq:Uratio}. Indeed, the expression for an arbitrary tesseral Love number
$k_{m+2,m}$ is given by 
\begin{align}\label{eq:km2m}
    k_{m+2,m}=
    &\frac{2\pi}{(2m+5)}
    \sum_{\alpha}
    \frac{Q_{m+2,m}^{\alpha}}
    {\epsilon_{\alpha}(\omega_{\alpha}-\omega_m)}
    \bigg\{
        Q_{m+2,m}^{\alpha}
        \\&\notag
        -\Omega_o^{-4/3}\left[ 
            \frac{(2m + 5)(4m+4)}
            {(2m + 1)^2}
        \right]^{1/2}Q_{mm}^{\alpha}
        +...
    \bigg\}.
\end{align}
The second term in curly brackets indicates a crucial dependence on the amplitude and sign of the ``cross-correlation'' $Q_{mm}^{\alpha}Q_{m+2,m}^{\alpha}.$ Numerically, we find that this cross-correlation is both negative and non-negligible for the sectoral f-modes $f_{mm}^\pm$, but negligible for most of the other modes (see \autoref{app:QvMode}). The driving of sectoral f-modes by the sectoral component of the tidal potential then produces the dominant $\propto\Omega_o^{-4/3}$ dependence (which derives from the ratio $U_{mm}/U_{m+2,m}$) illustrated for $k_{42},k_{53}$ and $k_{64}$ in \autoref{fig:n1poly_klm_modevar}. Even steeper power laws prevail at low $\Omega_o$ for larger $\ell>m+2$ (omitted for clarity), while $k_{31}$ does not show the same power law because the $m=1$ sectoral f-mode is a trivial mode with zero frequency.

This inverse dependence on the orbital frequency of the satellite 
\citep[similarly found in CMS calculations, and \emph{Juno} observations:][]{Wahl2017,Wahl2020,Nettelmann2019,Durante2020} 
implies that an infinitely distant satellite would produce an infinite tesseral Love number. However, this divergence does not have any physical consequence; it merely indicates that in a centrifugally distorted planet the response $\delta\Phi_{m+2,m}$ to a distant satellite with small $\Omega_o$ will be primarily driven by $U_{mm}$, since $|U_{mm}|\gg |U_{m+2,m}|$ for large separation. $\delta\Phi_{m+2,m}$ becomes large relative to $U_{m+2,m},$ but not in an absolute sense: $\delta\Phi_{m+2,m}/U_{mm}$ remains finite, and in fact $\delta\Phi_{m+2,m}$ decreases with increasing $a$. For very rapid rotation, it may be more useful to consider the full complement of potential Love numbers \citep[see, e.g., Eq. 37 in ][]{Ogilvie2013}, rather than just the subset relating the components of the response and the tidal potential with the same $\ell$ and $m$. 

\subsubsection{The effect of i-modes on Love numbers}
As illustrated in \autoref{fig:n1poly_klm_modevar} and \autoref{fig:n1poly_dklm_modevar}, the i-modes encounter resonances where their frequencies coincide with the tidal frequency. In the absence of dissipation, these resonances produce infinite peaks in the Love numbers; the peaks in \autoref{fig:n1poly_klm_modevar} and \autoref{fig:n1poly_dklm_modevar} have a finite height because of the finite grid-spacing in $\Omega_o$. Dissipation would introduce physically meaningful peak amplitudes, and could additionally broaden the resonances.

The resonances are narrow for the sectoral Love numbers $k_{mm},$ since the inertial modes couple weakly to the sectoral tides, and wider for the tesseral $k_{m+2,m}$. For instance, for $m=2$ the longest wavelength retrograde mode $i_{201}^-$ produces a relatively narrow resonance in $k_{22}$ at $\Omega_o/\Omega\simeq0.45,$ and a more significant dynamical correction to $k_{42}$ at the same frequency. The retrograde inertial modes $i_{301}^-$ and $i_{401}^-$ similarly affect the tesseral Love numbers $k_{53}$ and $k_{64}$ (respectively).

\begin{figure*}
    \centering
    \includegraphics[width=\columnwidth]{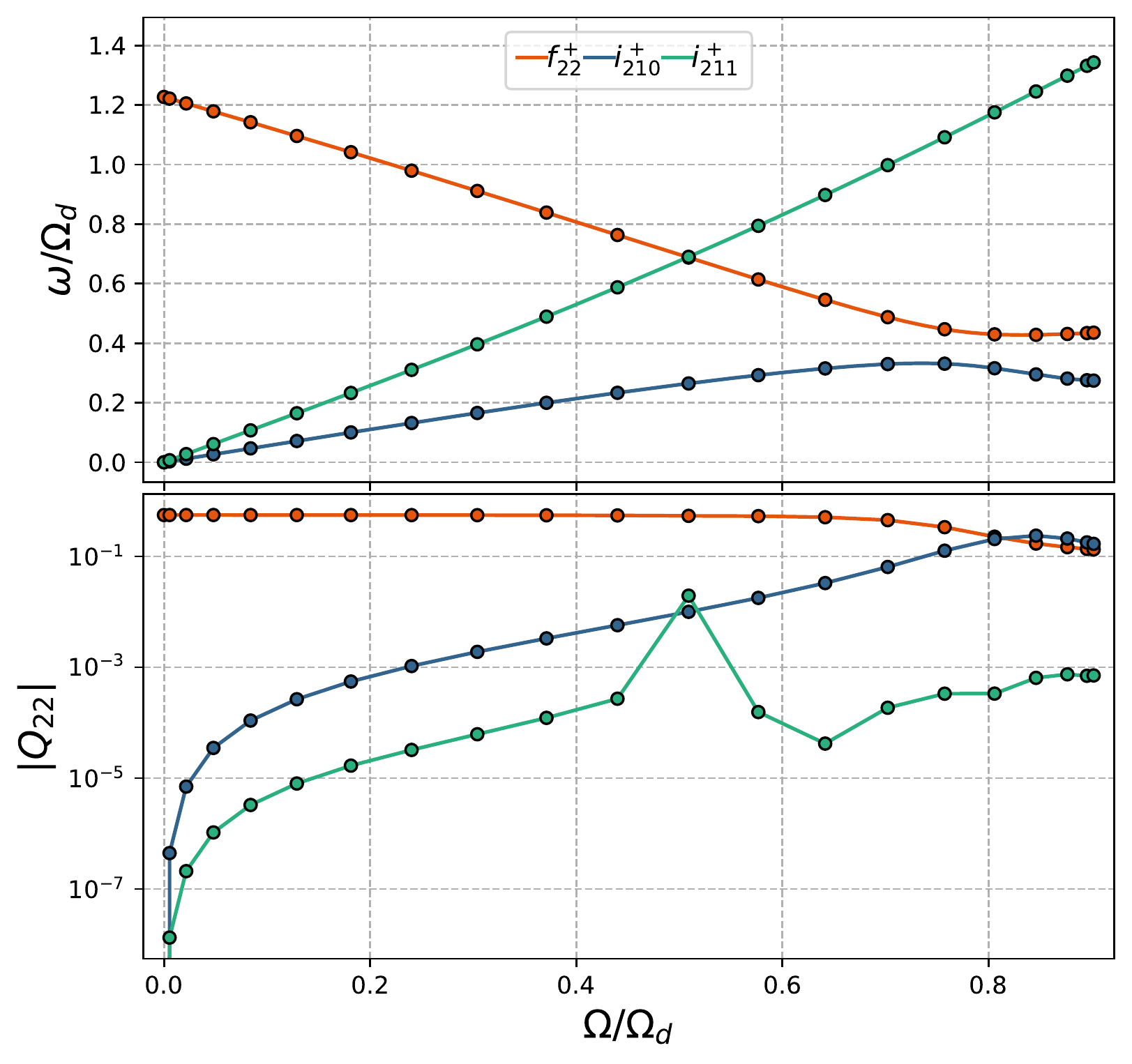}
    \includegraphics[width=\columnwidth]{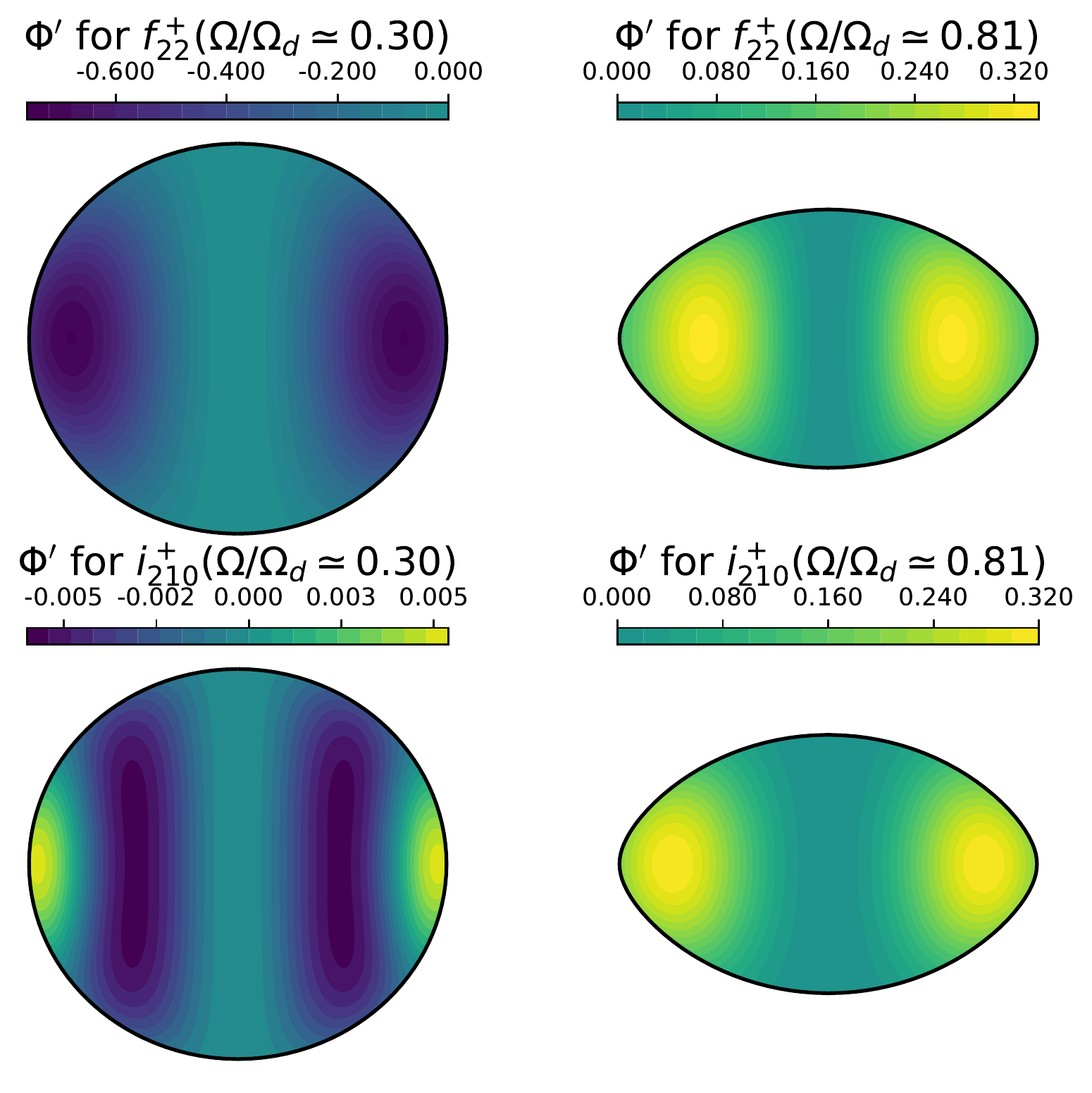}
    \caption{Left-hand panels: Frequencies (top) and tidal coupling coefficients $Q_{22}$ calculated for the sectoral $m=2$ f mode $f_{22}^+$ (orange) and the two prograde inertial modes $i_{210}^+$ and $i_{211}^+$ (blue and cyan). The mixing (``avoided crossing'') between $f_{22}^+$ and $i_{211}^+$ occurs over such a narrow frequency range that even the (indistinguishable) frequency separation at $\Omega/\Omega_d\simeq0.51$ is insufficient to amplify the i-mode's $|Q_{22}|$ beyond $0.1$. On the other hand, $f_{22}^+$ and $i_{210}^+$ experience strong frequency repulsion, and their wavefunctions overlap for an extended range of rotation rates $\Omega/\Omega_d\gtrsim0.7$. Right-hand colorplots: comparison cross-sections showing meridional slices of $\Phi'$ for $f_{22}^+$ (top) and $i_{210}^+$ (bottom) at rotation rates $\Omega/\Omega_d\simeq0.3$ (left) and $\Omega/\Omega_d\simeq0.81$ (right). At moderate rotation, the eigenfunction for $i_{210}^+$ is qualitatively distinct from that of $f_{22}^+$, and its gravitational perturbation is small under the normalization $\langle\boldsymbol{\xi},\boldsymbol{\xi}\rangle=1$. During the mixing around $\Omega/\Omega_d\gtrsim0.7$, however, the two oscillations lose distinction.}
    \label{fig:f2avcross} 
\end{figure*}

This stronger tesseral response of inertial modes was absent from the calculations of \citet{Lai2021}, who considered the effects of the longest wavelength $m=2$ i-modes on $k_{22}$ but not $k_{42}.$ The effect is interesting in light of the significant deviation between the $k_{42}$ inferred from \emph{Juno} observations, and the theoretical $k_{42}$ computed via the CMS method. We find $k_{42}\simeq1.54$ at Io's orbital frequency; discrepancies between this and the observed value of $k_{42}$  \citep[measured as $\simeq1.29$ by][]{Durante2020} could be explained by resonances involving retrograde i-modes with frequencies shifted by, e.g., a stably stratified interior.

The sub-inertial mode spectrum of course becomes more complicated for realistic planetary interior models including discrete regions of stable stratification, as the regular spectrum of pure inertial modes present in isentropic polytropes is replaced by mixed gravito-inertial modes \citep[e.g.,][]{Ouazzani2020}. However, even inertial waves confined to a convective envelope maintain a strong tesseral response \citep{Ogilvie2013}. We therefore expect the strong tesseral response of retrograde gravito-inertial modes close to resonance with the tidal forcing frequency to remain relevant in more realistic planetary models with stably stratified interiors.

\subsection{More rapid rotation}\label{sec:rapid}
In this subsection we describe our results for more rapid rotation ($\Omega/\Omega_d\simeq0.5-0.9$). While these results are primarily of academic interest when it comes to planetary science, they may bear relevance to very rapidly rotating neutron stars.

\subsubsection{Mixing between f-modes and i-modes}
As the rotation rate increases, the prograde f-modes $f_{22}^+$, $f_{33}^+$ and $f_{44}^+$ cross into the sub-inertial frequency range $\omega<2\Omega$ at $\Omega/\Omega_d\gtrsim0.4$ \citep[as found in MacLauren spheroids by ][]{Barker2016}. This leads to mode mixing or ``avoided crossings''---mode interactions in which oscillations come close to one another in frequency and trade physical character---between f-modes and i-modes. \autoref{fig:f2avcross} illustrates two such avoided crossings, between $f_{22}^+$ and both $i_{211}^+$ (at $\Omega/\Omega_d\simeq0.51$), and 
$i_{210}^+$ (at $\Omega/\Omega_d\simeq0.81$). The left panels show the frequencies $\omega$ (top) and overlap coefficient amplitudes $|Q_{22}|$ (bottom) as a function of $\Omega$ for all three modes. Meanwhile, the cross-sections on the right compare the gravitational perturbations for $f_{22}^+$ (top) and $i_{210}^+$ (bottom), for $\Omega/\Omega_d\simeq0.30$ (left) and for  $\Omega/\Omega_d\simeq0.81$ (right).

As shown in the top left panel of \autoref{fig:f2avcross}, the avoided crossing between $f_{22}^+$ and $i_{211}^+$ takes place over a narrow range of frequency. Consequently, the eigenfunction overlap associated with even the small frequency separation near $\Omega/\Omega_d\simeq0.51$ is insufficient to push the value of $|Q_{22}|$ for $i_{211}^+$ beyond $0.1.$ 

In contrast, the mixing between $f_{22}^+$ and $i_{210}^+$ involves strong frequency repulsion, such that the two oscillations are physically very similar for rotation rates $\Omega/\Omega_d\sim0.7-0.9$. The interaction imbues this particular prograde inertial mode with a strong coupling to the $\ell=m=2$ tidal potential, so that for these rapid rotation rates there are effectively two sectoral, prograde f-modes. As a result, it is important to include $i_{210}^+$ in Love number computations for $\Omega/\Omega_d\gtrsim0.7$. 

We note that \citet{Papaloizou2010} identified an avoided crossing between $i_{210}^+$ and another i-mode in their calculations, which excluded both centrifugal distortion and gravitational perturbations. We similarly find that i-modes with different wavelengths can undergo avoided crossings, but these are generally less important to the tidal response than the mixing between i-modes and f-modes demonstrated in \autoref{fig:f2avcross}.

\subsubsection{Love numbers for very rapid rotators}
\begin{figure*}
    \centering
    \includegraphics[width=.92\columnwidth]{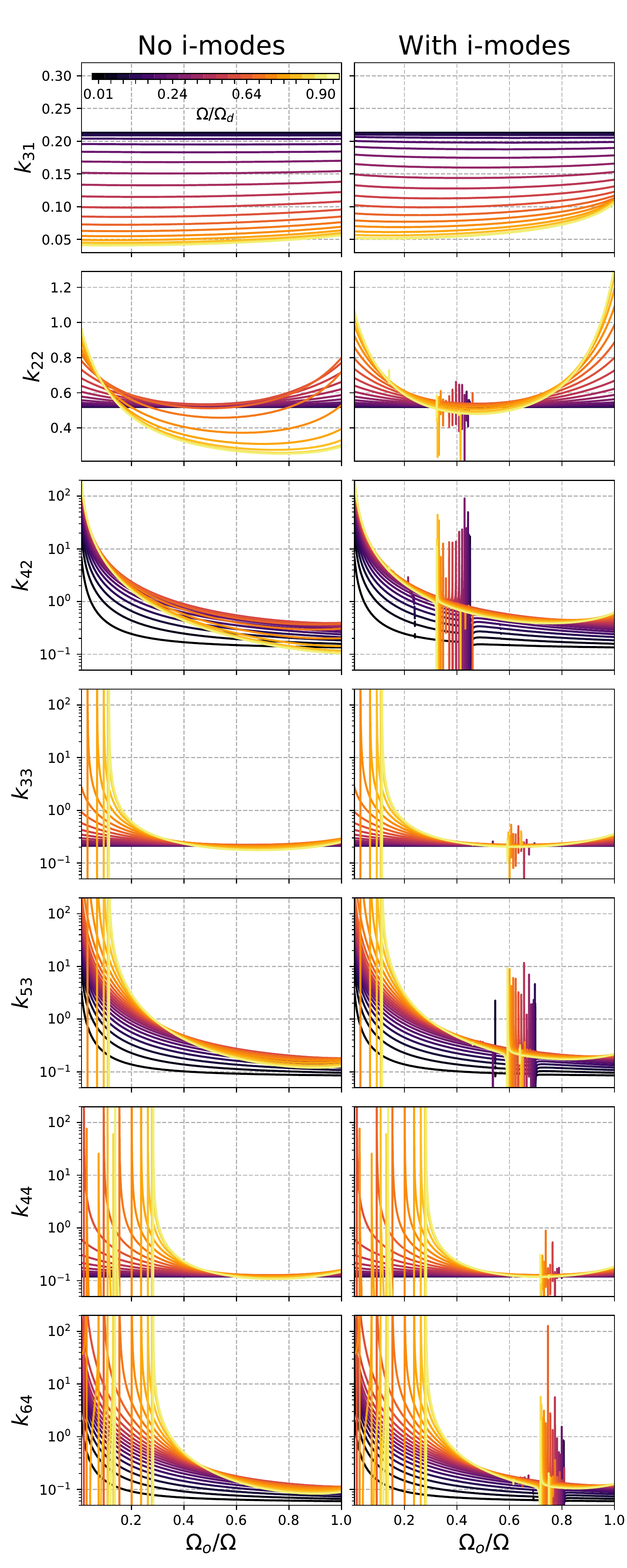}
    \includegraphics[width=.92\columnwidth]{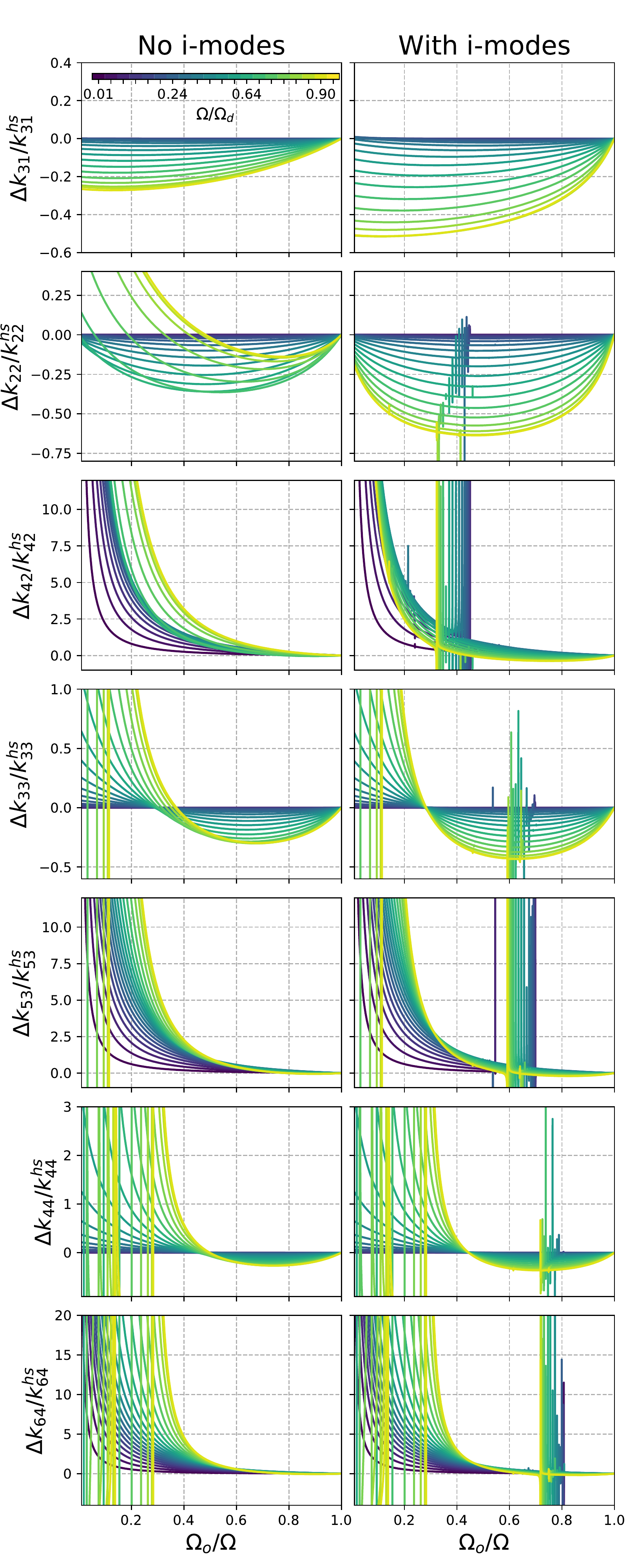}
    \caption{Love numbers (left panels) and dynamical Love number corrections (right panels), both with and without the inclusion of inertial modes (with $n_1+n_2\leq2$). The prograde inertial mode $i_{210}^+$ contributes more strongly to the hydrostatic Love number $k_{22}^\text{hs}=k_{22}(\Omega_o/\Omega=1)$ and $k_{42}^\text{hs}$ for rotation rates $\Omega\gtrsim0.7$, as it mixes with $f_{22}^+$ (see \autoref{fig:f2avcross}). In all but $k_{31}$, retrograde i-modes produce resonant features that are wider for the tesseral Love numbers with $\ell>m.$ The tesseral Love numbers additionally diverge as $\Omega_o\rightarrow0,$ due to contributions from sectoral f-modes driven by the sectoral tidal potential (see Section \ref{sec:fmneg}). On the other hand, for rapid rotation rates $k_{33}$, $k_{53}$, $k_{44}$ and $k_{64}$ exhibit more meaningful resonances at orbital frequencies $\Omega_o>0$, as the retrograde sectoral f-modes $f_{33}^-$ and $f_{44}^-$ pass into the CFS-unstable regime (this occurs when $\sigma=\omega+m\Omega=m\Omega_o$).
    }
    \label{fig:n1poly_klm_rotvar} 
\end{figure*}

\autoref{fig:n1poly_klm_rotvar} shows our Love number calculations at multiple rotation rates for an $n=1$ polytrope, as indicated by color schemes that extend from $\Omega/\Omega_d\simeq0.01$ (dark) to $\Omega/\Omega_d\simeq0.90$ (light). The left panel set shows $k_{\ell m}$ vs. $\Omega_o/\Omega$ (like \autoref{fig:n1poly_klm_modevar}), while the right panel set shows dynamical corrections to $k_{\ell m}^\text{hs}$ (like \autoref{fig:n1poly_dklm_modevar}). 

In both the left and right sets of panels, the separate columns show Love numbers computed with (right) and without (left) the contributions from the 
inertial modes included in this study
. Although we have checked that adding additional modes does not significantly alter the results summarized in \autoref{fig:n1poly_klm_rotvar}, for the most rapid rotation rates ($\Omega\gtrsim0.8\Omega_d$) centrifugal distortion is strong enough (see \autoref{fig:f2avcross}, right) that our truncated mode expansion should be treated with caution, since additional, shorter-wavelength modes might play a role in the aggregate. 

\autoref{fig:n1poly_klm_rotvar} demonstrates some qualitative differences between the tidal response with/without the inclusion of i-modes, in particular the longest wavelength $i_{m01}^-$ and $i_{m10}^+$. First, the strong overlap of $i_{201}^-$, $i_{301}^-$ and $i_{401}^-$ with the tesseral $(\ell m)=(42)$, $(53)$, and $(64)$ components of the tide persists at all rotation rates, as seen in the panels showing $k_{42}$, $k_{53}$, and $k_{64}$ (resp.). The tesseral coupling of $i_{101}^-$ with the $(\ell m)=(31)$ tidal potential also affects $k_{31}$, but the mode frequency is too negative to resonate with the tidal forcing frequencies considered. As in the case of $\Omega/\Omega_d\simeq0.3$, the shorter wavelength, retrograde i-modes with $n_1+n_2=2$ contribute much narrower resonances to all the other Love numbers. 

Secondly, the avoided crossing between the prograde modes $f_{22}^+$ and $i_{210}^+$ (see \autoref{fig:f2avcross}) results in significant differences between the $k_{22}$ and $k_{42}$ values calculated with and without the inclusion of the latter mode, even far from resonance (see the $\Delta k_{22}$ and $\Delta k_{42}$ panels). This considerably alters the zero-frequency response ($k_{\ell m}^\text{hs}$) with/without i-modes. For very rapid rotation ($\Omega/\Omega_d\gtrsim0.87$), mixing with the $m=2$ f-mode causes even the shorter-wavelength, prograde i-mode $i_{220}^+$ to have a larger impact on the hydrostatic Love numbers; we find that excluding it leads to a decrease in $k_{22}^\text{hs}$ and $k_{42}^\text{hs}$.

At all rotation rates, forcing of the sectoral f-modes by the $\ell=m$ component of the tide, along with non-zero cross-correlations $Q_{mm}Q_{m+2,m}$ due to centrifugal distortion, leads to a divergence in the tesseral Love numbers ($k_{42}$, $k_{53}$, and $k_{64}$) as $\Omega_o\rightarrow0.$ However, as discussed in Section \ref{sec:fmneg}, this divergence is not particularly meaningful in that $\delta \Phi_{42}$, $\delta \Phi_{53}$, and $\delta \Phi_{64}$ remain finite (for moderate rotation rates), only becoming large relative to $U_{42}$, $U_{53}$, and $U_{64}$ (respectively).

On the other hand, at very rapid rotation rates $k_{33}$, $k_{53}$, $k_{44}$ and $k_{64}$ exhibit more meaningful divergence at orbital frequencies $\Omega_o/\Omega>0,$ due to resonances: as indicated by \autoref{fig:n1poly_omQlm} (top left), the frequencies of the f-modes $f_{33}^-$ and $f_{44}^-$ (negative under our convention) decrease less rapidly than $-m\Omega$. For sufficiently rapid rotation rates ($\Omega/\Omega_d\gtrsim0.8$), this means that the resonant response of these oscillations falls into the plotted orbital frequency range $\Omega_o/\Omega\in[0,1]$.

These resonances arise because f-modes that are retrograde in the rotating frame ($\omega<0$) can become prograde in the inertial frame for rotation rates large enough that their inertial-frame frequency is positive ($\sigma=\omega+m\Omega>0$). We note that the property of propagating backward in the rotating frame, but forward in an inertial frame renders these modes unstable to secular excitation via  the Chandrasekhar-Friedman-Schutz (CFS) instability \citep{Chandrasekhar1970,Friedman1978}. The CFS instability involves
the enhancement of retrograde modes' negative angular momentum, due to the removal of positive angular momentum by gravitational wave radiation.

The required rotation rates for the onset of the CFS instability for sectoral f-modes are quite large for small $m$, and decrease for larger $m$. \citet{Ipser1990} showed that for an $n=1$ polytrope, all the retrograde sectoral f-modes save for $f_{22}^-$ can experience CFS instabiity before the mass-shedding (break-up) rotation limit.  In any case, the required rotaton rates are unlikely to be realized in giant planets. Such rapid rotation is not excluded for neutron stars, however; as found by \citet{Ho1999}, resonant f-mode excitation in inspiralling neutron star binaries may affect the observed gravitational waveform. 

\section{Summary and conclusions}\label{sec:conc}
We have developed a new, precise method for computing the dynamical, non-dissipative tidal response of rapidly rotating planets and stars, modeled in this paper by isentropic polytropes.  The method involves summing over contributions to the potential Love numbers $k_{\ell m}$ from normal mode oscillations, which we have computed using a non-perturbative treatment of rotation that fully accounts for the effects of both the Coriolis force and centrifugal distortion. 

\subsection{Moderate rotation rates}
Focusing on the polytropic index $n=1$ most relevant to giant planets and neutron stars, we have evaluated the relative importance of fundamental modes (f-modes), acoustic modes (p-modes), and inertial modes (i-modes), as dictated by their frequencies $\omega$ and tidal overlap coefficients $Q_{\ell m}$ (see \autoref{fig:n1pOm3Q}). Identifying the most significant modes for rotation rates $\Omega\lesssim0.4\Omega_d$, where $\Omega_d=(GM/R_\text{eq}^3)^{1/2}$ is the breakup rotation rate, we have provided analytical expressions (polynomial fits; see \autoref{fig:n1poly_omQlm}; Table \ref{tab:n1fmode}; Table \ref{tab:n1imode}) to the quantities required to compute a variety of Love numbers. 

For most rotation rates in this range, a perturbative treatment of rotation (to linear order in $\Omega$) focused on f-modes is adequate for capturing the sectoral ($\ell=|m|$) dynamical tidal Love numbers, to which p-modes and i-modes contribute negligibly \citep{Lai2021}. Such a perturbative calculation can, for instance, satisfactorily explain the discrepancy between the value of Jupiter's $k_{22}$  inferred from \emph{Juno} observations, and predicted by calculations ignoring dynamical effects. However, the tesseral ($\ell>|m|$) coupling of retrograde i-modes to the tidal potential is more important to consider than their sectoral coupling; we show that the ``leading'' $m=2$ inertial modes can induce broad resonant features in (e.g.) $k_{42}$, significantly modifying the dynamical corrections to the hydrostatic value (see Figures \ref{fig:n1poly_klm_modevar}-\ref{fig:n1poly_dklm_modevar}). 

Additionally, we show that sectoral f-modes driven by the sectoral components of the tidal potential ($U_{\ell m}$ with $\ell=m$) can produce a finite tesseral gravitational response ($\delta\Phi_{\ell m},$ with $\ell>m$)---an order ${\cal O}(\Omega^2)$ effect that is not captured by perturbative treatments of rotation that ignore centrifugal distortion of the planet or star. Driving of the tesseral tidal response by sectoral components of the tidal potential results in tesseral Love numbers that diverge with increasing companion separation (see Section \ref{sec:fmneg}). This divergence suggests that alternative response functions (e.g., potential Love numbers relating unlike components of the tidal potential and the response) may be more useful for characterizing the response of significantly centrifugally distorted bodies.

\subsection{Rapid rotation rates}
We have also considered the tidal response of much more rapid rotators. For $\Omega/\Omega_d\gtrsim0.7$, mixing between prograde, sectoral f-modes and prograde i-modes can enhance the overlap of the latter with the sectoral components of the tidal potential (\autoref{fig:f2avcross}; panels showing $k_{22}$ in  \autoref{fig:n1poly_klm_rotvar}). These ``avoided crossings'' make the contributions of prograde i-modes important to consider when characterizing the tidal response of rapidly rotating planets or stars. 

Lastly, the criterion of negative (positive) frequencies in the rotating frame (inertial frame) for secular CFS instability also demarcates a regime in which retrograde, sectoral f-modes with $m>2$ may come into resonance with tidal forcing frequencies associated with distant, spin-orbit aligned companions (bottom four panels of \autoref{fig:n1poly_klm_rotvar}). Such resonant f-modes may significantly alter the dynamical corrections to both the sectoral and tesseral Love numbers $k_{mm}$ and $k_{m+2,m}$ in very rapidly rotating, centrifugally distorted bodies.

\begin{acknowledgments}
We thank the anonymous referee who reviewed this manuscript, and provided thorough and constructive comments that significantly improved the quality of the paper. J. W. D. gratefully acknowledges support from the Natural Sciences and Engineering Research Council of Canada (NSERC) [funding reference $\#$CITA 490888-16], and from the Sloan Foundation through grant FG-2018-10515.
\end{acknowledgments}

%



\bibliography{polytables}{}

\begin{thebibliography}{}
\expandafter\ifx\csname natexlab\endcsname\relax\def\natexlab#1{#1}\fi
\providecommand{\url}[1]{\href{#1}{#1}}
\providecommand{\dodoi}[1]{doi:~\href{http://doi.org/#1}{\nolinkurl{#1}}}
\providecommand{\doeprint}[1]{\href{http://ascl.net/#1}{\nolinkurl{http://ascl.net/#1}}}
\providecommand{\doarXiv}[1]{\href{https://arxiv.org/abs/#1}{\nolinkurl{https://arxiv.org/abs/#1}}}

\bibitem[{{Barker} {et~al.}(2016){Barker}, {Braviner}, \&
  {Ogilvie}}]{Barker2016}
{Barker}, A.~J., {Braviner}, H.~J., \& {Ogilvie}, G.~I. 2016, \mnras, 459, 924

\bibitem[{{Bonazzola} {et~al.}(1998){Bonazzola}, {Gourgoulhon}, \&
  {Marck}}]{Bonazzola1998}
{Bonazzola}, S., {Gourgoulhon}, E., \& {Marck}, J.-A. 1998, \prd, 58, 104020

\bibitem[{{Boyd}(2001)}]{Boyd2001}
{Boyd}, J.~P. 2001, {Chebyshev and Fourier Spectral Methods} (Dover
  Publications, Inc)

\bibitem[{{Boyd}(2011)}]{Boyd2011}
---. 2011, Numerical Mathematics: Theory, Methods and Applications, 4, 142

\bibitem[{{Braviner} \& {Ogilvie}(2014)}]{Braviner2014}
{Braviner}, H.~J., \& {Ogilvie}, G.~I. 2014, \mnras, 441, 2321

\bibitem[{{Braviner} \& {Ogilvie}(2015)}]{Braviner2015}
---. 2015, \mnras, 447, 1141

\bibitem[{{Bryan}(1889)}]{Bryan1889}
{Bryan}, G.~H. 1889, Philosophical Transactions of the Royal Society of London
  Series A, 180, 187

\bibitem[{{Chandrasekhar}(1970)}]{Chandrasekhar1970}
{Chandrasekhar}, S. 1970, \prl, 24, 611

\bibitem[{{De Pietri} {et~al.}(2018){De Pietri}, {Feo}, {Font}, {L{\"o}ffler},
  {Maione}, {Pasquali}, \& {Stergioulas}}]{dePietri2018}
{De Pietri}, R., {Feo}, A., {Font}, J.~A., {et~al.} 2018, \prl, 120, 221101

\bibitem[{{Dewberry} {et~al.}(2021){Dewberry}, {Mankovich}, {Fuller}, {Lai}, \&
  {Xu}}]{Dewberry2021}
{Dewberry}, J.~W., {Mankovich}, C.~R., {Fuller}, J., {Lai}, D., \& {Xu}, W.
  2021, \psj, 2, 198

\bibitem[{{Durante} {et~al.}(2020){Durante}, {Parisi}, {Serra}, {Zannoni},
  {Notaro}, {Racioppa}, {Buccino}, {Lari}, {Gomez Casajus}, {Iess}, {Folkner},
  {Tommei}, {Tortora}, \& {Bolton}}]{Durante2020}
{Durante}, D., {Parisi}, M., {Serra}, D., {et~al.} 2020, \grl, 47, e86572

\bibitem[{{Friedman} \& {Schutz}(1978)}]{Friedman1978}
{Friedman}, J.~L., \& {Schutz}, B.~F. 1978, \apj, 222, 281

\bibitem[{{Goodman} \& {Lackner}(2009)}]{Goodman2009}
{Goodman}, J., \& {Lackner}, C. 2009, \apj, 696, 2054

\bibitem[{{Greenspan}(1968)}]{Greenspan1968}
{Greenspan}, H.~P. 1968, {The Theory of Rotating Fluids} (Cambridge Univ.
  Press)

\bibitem[{{Hachisu}(1986)}]{Hachisu1986}
{Hachisu}, I. 1986, \apjs, 62, 461

\bibitem[{{Ho} \& {Lai}(1999)}]{Ho1999}
{Ho}, W. C.~G., \& {Lai}, D. 1999, \mnras, 308, 153

\bibitem[{{Idini} \& {Stevenson}(2021)}]{Idini2021}
{Idini}, B., \& {Stevenson}, D.~J. 2021, \psj, 2, 69

\bibitem[{{Ipser} \& {Lindblom}(1990)}]{Ipser1990}
{Ipser}, J.~R., \& {Lindblom}, L. 1990, \apj, 355, 226

\bibitem[{{Ivanov} \& {Papaloizou}(2010)}]{Ivanov2010}
{Ivanov}, P.~B., \& {Papaloizou}, J.~C.~B. 2010, \mnras, 407, 1609

\bibitem[{{Jackson}(1962)}]{Jackson1962}
{Jackson}, J.~D. 1962, {Classical Electrodynamics} (Wiley)

\bibitem[{{Lai}(2021)}]{Lai2021}
{Lai}, D. 2021, \psj, 2, 122

\bibitem[{{Lai} \& {Wu}(2006)}]{Lai2006}
{Lai}, D., \& {Wu}, Y. 2006, \prd, 74, 024007

\bibitem[{{Lainey} {et~al.}(2009){Lainey}, {Arlot}, {Karatekin}, \& {van
  Hoolst}}]{Lainey2009}
{Lainey}, V., {Arlot}, J.-E., {Karatekin}, {\"O}., \& {van Hoolst}, T. 2009,
  \nat, 459, 957

\bibitem[{{Lainey} {et~al.}(2012){Lainey}, {Karatekin}, {Desmars}, {Charnoz},
  {Arlot}, {Emelyanov}, {Le Poncin-Lafitte}, {Mathis}, {Remus}, {Tobie}, \&
  {Zahn}}]{Lainey2012}
{Lainey}, V., {Karatekin}, {\"O}., {Desmars}, J., {et~al.} 2012, \apj, 752, 14

\bibitem[{{Lainey} {et~al.}(2017){Lainey}, {Jacobson}, {Tajeddine}, {Cooper},
  {Murray}, {Robert}, {Tobie}, {Guillot}, {Mathis}, {Remus}, {Desmars},
  {Arlot}, {De Cuyper}, {Dehant}, {Pascu}, {Thuillot}, {Le Poncin-Lafitte}, \&
  {Zahn}}]{Lainey2017}
{Lainey}, V., {Jacobson}, R.~A., {Tajeddine}, R., {et~al.} 2017, \icarus, 281,
  286

\bibitem[{{Lainey} {et~al.}(2020){Lainey}, {Casajus}, {Fuller}, {Zannoni},
  {Tortora}, {Cooper}, {Murray}, {Modenini}, {Park}, {Robert}, \&
  {Zhang}}]{Lainey2020}
{Lainey}, V., {Casajus}, L.~G., {Fuller}, J., {et~al.} 2020, Nature Astronomy,
  4, 1053

\bibitem[{{Lattimer} \& {Prakash}(2007)}]{Lattimer2007}
{Lattimer}, J.~M., \& {Prakash}, M. 2007, \physrep, 442, 109

\bibitem[{{Ligni{\`e}res} \& {Georgeot}(2009)}]{Lignieres2009}
{Ligni{\`e}res}, F., \& {Georgeot}, B. 2009, \aap, 500, 1173

\bibitem[{{Ligni{\`e}res} {et~al.}(2006){Ligni{\`e}res}, {Rieutord}, \&
  {Reese}}]{Lignieres2006}
{Ligni{\`e}res}, F., {Rieutord}, M., \& {Reese}, D. 2006, \aap, 455, 607

\bibitem[{{Lin} \& {Ogilvie}(2021)}]{Lin2021}
{Lin}, Y., \& {Ogilvie}, G.~I. 2021, \apjl, 918, L21

\bibitem[{{Lindblom} \& {Ipser}(1999)}]{Lindblom1999}
{Lindblom}, L., \& {Ipser}, J.~R. 1999, \prd, 59, 044009

\bibitem[{{Lockitch} \& {Friedman}(1999)}]{Lockitch1999}
{Lockitch}, K.~H., \& {Friedman}, J.~L. 1999, \apj, 521, 764

\bibitem[{{Mankovich} \& {Fuller}(2021)}]{Mankovich2021}
{Mankovich}, C., \& {Fuller}, J. 2021, NatAs,
  \dodoi{10.1038/s41550-021-01448-3}

\bibitem[{{Nettelmann}(2019)}]{Nettelmann2019}
{Nettelmann}, N. 2019, \apj, 874, 156

\bibitem[{{Ogilvie}(2005)}]{Ogilvie2005}
{Ogilvie}, G.~I. 2005, Journal of Fluid Mechanics, 543, 19

\bibitem[{{Ogilvie}(2009)}]{Ogilvie2009}
---. 2009, \mnras, 396, 794

\bibitem[{{Ogilvie}(2013)}]{Ogilvie2013}
---. 2013, \mnras, 429, 613

\bibitem[{{Ogilvie}(2014)}]{Ogilvie2014}
---. 2014, \araa, 52, 171

\bibitem[{{Ouazzani} {et~al.}(2012){Ouazzani}, {Dupret}, \&
  {Reese}}]{Ouazzani2012}
{Ouazzani}, R.~M., {Dupret}, M.~A., \& {Reese}, D.~R. 2012, \aap, 547, A75

\bibitem[{{Ouazzani} {et~al.}(2020){Ouazzani}, {Ligni{\`e}res}, {Dupret},
  {Salmon}, {Ballot}, {Christophe}, \& {Takata}}]{Ouazzani2020}
{Ouazzani}, R.~M., {Ligni{\`e}res}, F., {Dupret}, M.~A., {et~al.} 2020, \aap,
  640, A49

\bibitem[{{Papaloizou} \& {Ivanov}(2010)}]{Papaloizou2010}
{Papaloizou}, J.~C.~B., \& {Ivanov}, P.~B. 2010, \mnras, 407, 1631

\bibitem[{{Passamonti} {et~al.}(2009{\natexlab{a}}){Passamonti}, {Haskell}, \&
  {Andersson}}]{Passamonti2009b}
{Passamonti}, A., {Haskell}, B., \& {Andersson}, N. 2009{\natexlab{a}}, \mnras,
  396, 951

\bibitem[{{Passamonti} {et~al.}(2009{\natexlab{b}}){Passamonti}, {Haskell},
  {Andersson}, {Jones}, \& {Hawke}}]{Passamonti2009a}
{Passamonti}, A., {Haskell}, B., {Andersson}, N., {Jones}, D.~I., \& {Hawke},
  I. 2009{\natexlab{b}}, \mnras, 394, 730

\bibitem[{{Press} \& {Teukolsky}(1977)}]{Press1977}
{Press}, W.~H., \& {Teukolsky}, S.~A. 1977, \apj, 213, 183

\bibitem[{{Reese} {et~al.}(2006){Reese}, {Ligni{\`e}res}, \&
  {Rieutord}}]{Reese2006}
{Reese}, D., {Ligni{\`e}res}, F., \& {Rieutord}, M. 2006, \aap, 455, 621

\bibitem[{{Reese} {et~al.}(2009){Reese}, {MacGregor}, {Jackson}, {Skumanich},
  \& {Metcalfe}}]{Reese2009}
{Reese}, D.~R., {MacGregor}, K.~B., {Jackson}, S., {Skumanich}, A., \&
  {Metcalfe}, T.~S. 2009, \aap, 506, 189

\bibitem[{{Reese} {et~al.}(2021){Reese}, {Mirouh}, {Espinosa Lara}, {Rieutord},
  \& {Putigny}}]{Reese2021}
{Reese}, D.~R., {Mirouh}, G.~M., {Espinosa Lara}, F., {Rieutord}, M., \&
  {Putigny}, B. 2021, \aap, 645, A46

\bibitem[{{Reese} {et~al.}(2013){Reese}, {Prat}, {Barban}, {van 't
  Veer-Menneret}, \& {MacGregor}}]{Reese2013}
{Reese}, D.~R., {Prat}, V., {Barban}, C., {van 't Veer-Menneret}, C., \&
  {MacGregor}, K.~B. 2013, \aap, 550, A77

\bibitem[{{Rieutord} \& {Valdettaro}(1997)}]{Rieutord1997}
{Rieutord}, M., \& {Valdettaro}, L. 1997, JFM, 341

\bibitem[{{Rieutord} \& {Valdettaro}(2010)}]{Rieutord2010}
---. 2010, Journal of Fluid Mechanics, 643, 363

\bibitem[{{Schenk} {et~al.}(2002){Schenk}, {Arras}, {Flanagan}, {Teukolsky}, \&
  {Wasserman}}]{Schenk2002}
{Schenk}, A.~K., {Arras}, P., {Flanagan}, {\'E}.~{\'E}., {Teukolsky}, S.~A., \&
  {Wasserman}, I. 2002, \prd, 65, 024001

\bibitem[{{Unno} {et~al.}(1989){Unno}, {Osaki}, {Ando}, {Saio}, \&
  {Shibahashi}}]{Unno1989}
{Unno}, W., {Osaki}, Y., {Ando}, H., {Saio}, H., \& {Shibahashi}, H. 1989,
  {Nonradial oscillations of stars} (University of Tokyo Press)

\bibitem[{{Wahl} {et~al.}(2017){Wahl}, {Hubbard}, \& {Militzer}}]{Wahl2017}
{Wahl}, S.~M., {Hubbard}, W.~B., \& {Militzer}, B. 2017, \icarus, 282, 183

\bibitem[{{Wahl} {et~al.}(2020){Wahl}, {Parisi}, {Folkner}, {Hubbard}, \&
  {Militzer}}]{Wahl2020}
{Wahl}, S.~M., {Parisi}, M., {Folkner}, W.~M., {Hubbard}, W.~B., \& {Militzer},
  B. 2020, \apj, 891, 42

\bibitem[{{Wu}(2005)}]{Wu2005a}
{Wu}, Y. 2005, \apj, 635, 674

\bibitem[{{Xu} \& {Lai}(2017)}]{Xu2017}
{Xu}, W., \& {Lai}, D. 2017, \prd, 96, 083005

\end{thebibliography}
\bibliographystyle{aasjournal}





\appendix
\section{Oblate polytropic models}\label{app:modelcalc}
In the Newtonian limit, the equations governing the mechanical equilibrium of a rigidly rotating, self-gravitating, barotropic fluid body characterized by density $\rho$, pressure $P=P(\rho)$, gravitational field $\Phi$ and angular velocity $\boldsymbol{\Omega}=\Omega\hat{\bf z}$ are
\begin{align}\label{eq:Hstat}
   \frac{1}{\rho}\nabla P
   &=-\nabla\left(\Phi-\frac{1}{2}\Omega^2R^2\right),
\\
    \nabla^2\Phi&=4\pi G\rho,
\end{align}
where $G$ is the gravitational constant, and $R$ the cylindrical radius. Assuming a polytropic equation of state $P\propto\rho^{1+1/n}$, and defining a pseudo-enthalpy $H =\int \text{d} P /\rho =(1+n)P /\rho,$ \autoref{eq:Hstat} can be integrated directly to give
\begin{align}
   \frac{H }{H_c}=\left(\frac{\rho}{\rho_c}\right)^{1/n}
   &=1-\frac{1}{H_c}\left(
        \Phi -\Phi_c
        -\frac{1}{2}\Omega^2R^2
    \right),
\end{align}
where subscript $c$'s denote central values. $H_c$ can be fixed with the boundary condition $H=0$ on the surface $r=r_s(\mu)$ \citep{Hachisu1986}, where $\mu=\cos\theta$ is a convenient variable to use instead of colatitude. This boundary condition implies
\begin{align}
   H_c&=\Phi_\text{eq}-\Phi_c-\frac{1}{2}\Omega^2R_\text{eq}^2
   =\Phi_\text{pol}-\Phi_c,
\end{align}
where $\Phi_\text{eq}=\Phi(r=R_\text{eq},\mu=0)$ and $\Phi_\text{pol}=\Phi(r=R_\text{pol},\mu=1)$ are evaluated at the equator and poles, respectively. We non-dimensionalize by writing $r=R_\text{eq}\tilde{r}$, $\Omega=\tilde{\Omega}R_\text{eq}^{-1}\sqrt{H_c}$, and immediately suppressing tildes.

\subsection{Non-rotating polytropes}
In the absence of rotation, substition into Poisson's equation, together with the definition $\Theta=(\Phi_\text{pol}-\Phi)/(\Phi_\text{pol}-\Phi_c)$ leads to the Lane-Emden equation $\nabla^2\Theta=-\lambda\Theta^n,$ where $\lambda=4\pi G\rho_cR_\text{eq}^2/H_c$ can be treated as an eigenvalue. Assuming spherical symmetry, the appropriate radial boundary conditions are $\Theta'(0)=0,$ $\Theta(1)=0$, and $\Theta(0)=1$ (the third boundary condition for this second-order differential equation results from the choice to cast the radial scale in terms of an eigenvalue $\lambda,$ rather than solve for the first zero of $\Theta$). We solve for 1D, non-rotating polytropic models using Newton-Kantorovich iteration as described by \citet{Boyd2011}.

\subsection{Rotating polytropes}
Starting from spherically symmetric, non-rotating solutions, we solve for rapidly rotating polytropic models using an iterative scheme. Continuing to scale lengths by $R_\text{eq}$ and angular velocity by $R_\text{eq}^{-1}\sqrt{H_c}$, we additionally work with $\Phi$ in units of $H_c$ and $\rho$ in units of $\rho_c.$ The iterative scheme for calculating the ${k+1}^\text{th}$ iteration from the $k^\text{th}$ iteration can then be summarized as 
\begin{align}\label{eq:ischeme1}
    \nabla^2\Phi^{k+1}&=\lambda^k\rho^k,
\\\label{eq:ischeme2}
    H_c^{k+1}&=\Phi_\text{pol}^{k+1}-\Phi_c^{k+1},
\\
    \rho^{k+1}&=\left\{
        1 - (1/H_c^{k+1})
        \left[
            \Phi^{k+1}-\Phi_c^{k+1}-(1/2)\Omega^2R^2
        \right]
    \right\}^n,
\\\label{eq:ischeme4}
    \lambda^{k+1}&=\lambda^k/H_c^{k+1}.
\end{align}
Although we perform mode calculations in a non-orthogonal coordinate system constructed to match the oblate surface of the polytrope, since the surface radius is not known a priori we calculate models simply in spherical coordinates: we first expand the gravitational potential (and similarly the density profile) as 
\begin{align}
    \Phi(r,\mu)&=\sum_{\ell'=0}^{N_\ell}\Phi^{2\ell'}(r)Y_{2\ell'}^{m=0}(\mu).
\end{align}
Projected onto an arbitrary $Y_\ell^{m=0}$ of degree $\ell$, Poisson's equation for coefficients $\Phi^\ell$ and $\rho^\ell$ reads
\begin{equation}
    \frac{1}{r^2}\frac{\partial }{\partial r}\left(r^2\frac{\partial \Phi^\ell}{\partial r}\right)
    -\frac{\ell(\ell + 1)}{r^2}\Phi^\ell
    =\lambda \rho^\ell.
\end{equation}
We solve this equation in the radial domain $r/R_\text{eq}\in[0,2]$, using Chebyshev collocation with two radial domains split between $r/R_\text{eq}\in(0,1]$ and $[1,2].$ We enforce the regularity of each $\Phi^\ell$ at the origin, and the continuity of each $\Phi^\ell$ and its radial derivative at $r/R_\text{eq}=1$. At the outer boundary $r/R_\text{eq}=2$, we impose the boundary condition $[\partial_r+(\ell+1)/r]\Phi^\ell=0.$ We then reconstruct the 2D field $\Phi(r,\mu)$ from the coefficient $\Phi^\ell(r)$, before calculating subsequent values of $H_c$, $\rho$, and $\lambda$  from the algebraic Equations \ref{eq:ischeme2}-\ref{eq:ischeme4} [setting $\rho=0$ exterior to the surface $r_s(\mu)$ defined by $H=0$].

\begin{figure*}
    \centering
    \includegraphics[width=\textwidth]{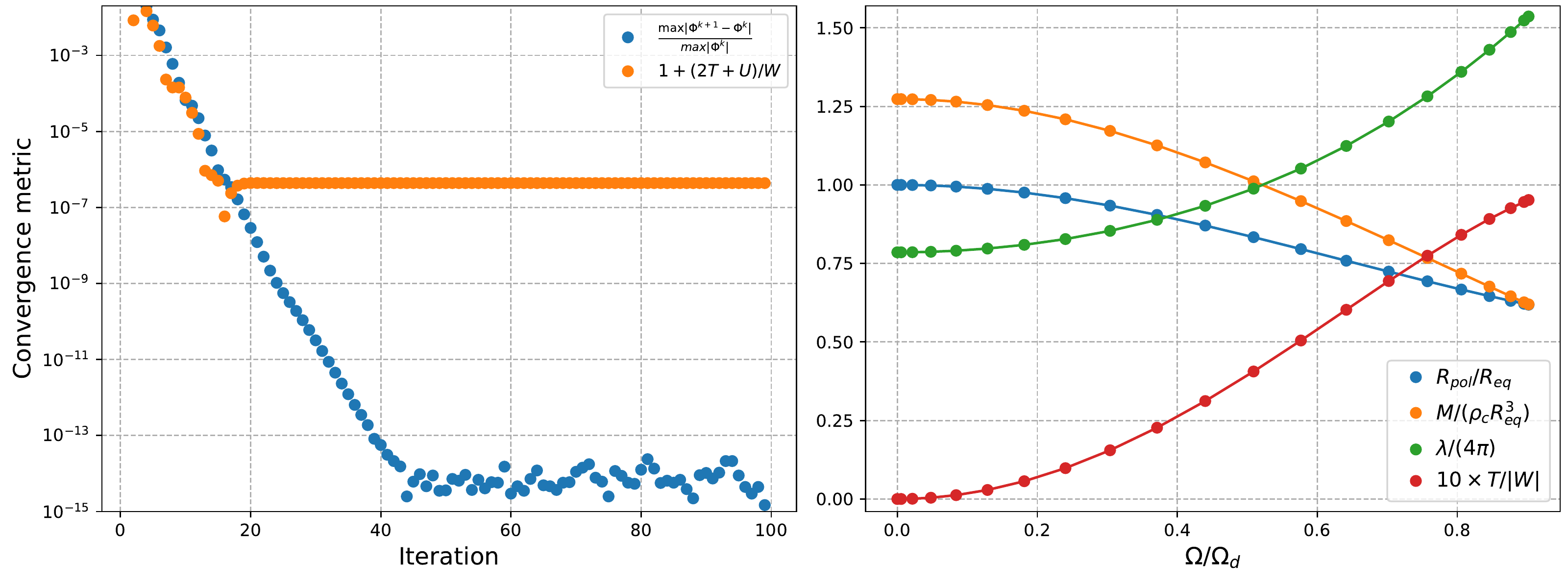}
    \caption{Left: relative changes in $\Phi(r,\mu)$ (blue) and virial errors $\epsilon_V$ (orange) with successive iterations for calculations of a near-maximally rotating, $n=1$ polytrope with $\Omega/\Omega_d\simeq0.99$  (note that less rapidly rotating models converge with fewer iterations). Right: variation of model parameters for the $n=1$ polytropic models considered in this paper. 
    }
    \label{fig:n1poly}
\end{figure*}
\autoref{fig:n1poly} (left) illustrates the convergence of the iteration scheme for a model with polytropic index $n=1$ and nearly maximal $\Omega/\Omega_d\simeq0.99$. The plot shows relative changes in gravitational field with each iteration (blue), and also a virial error defined by $\epsilon_V=1+(2T+U)/W$ (orange), where
\begin{equation}\label{eq:modelE}
    T=\frac{1}{2}\int_V\rho\Omega^2R^2\text{d}V,
    \hspace{4em}
    U=3\int_VP\text{d}V,
    \hspace{4em}
    W=\frac{1}{2}\int_V\rho\Phi\text{d}V.
\end{equation}
The virial error assesses the degree to which the stellar model satisfies the virial theorem $2T+U+W=0;$ we achieve $\epsilon_v\lesssim10^{-6}$ for all models considered in this paper (with much smaller errors for more centrally condensed polytropes). \autoref{fig:n1poly} (right) plots the ratio between polar and equatorial radius (blue), total mass (orange), eigenvalue $\lambda/(4\pi)=G\rho_cR_\text{eq}^2/H_c$ (green), and ten times the ratio of kinetic to gravitaional potential energy, $10T/|W|$. Accounting for differences in dimensionalization, we find values in agreement with \citet{Reese2006} and \citet{Passamonti2009a}. 

\section{Numerical details}\label{app:mres}
The mode (model) calculations presented in this work were performed with $N_\zeta=200$ ($N_r=400$) Chebyshev collocation points $\zeta=-x\in(0,1]$, where $x=\cos[\pi i/(2N_\zeta-1)],$ $i=N_\zeta,...,2N_\zeta-1$ constitute half of a Gauss-Lobatto (endpoint-extrema) grid covering the stellar/planetary interior. We have halved the grid and constructed spectral derivative matrices from exclusively even or odd Chebyshev polynomials in order to apply regularity boundary conditions at $\zeta=0$ implicitly \citep[see Chapter 8 in ][]{Boyd2001}, but we note that we recover the same results using a full Gauss-Lobatto grid and explicit boundary condition enforcement via boundary bordering. In the exterior vacuum, our mode calculations used a (full) Gauss-Lobatto grid with $N_\zeta=31$, and spherical harmonic expansions varying from $N_\ell=8$ to $N_\ell=50$ (depending on rotation rate). Both frequencies and eigenfunctions were validated against calculations at lower resolution (in both $N_\zeta$ and $N_\ell$).

\section{Tidal overlap coefficients}\label{app:QvMode}
\begin{figure*}
    \centering
    \includegraphics[width=\textwidth]{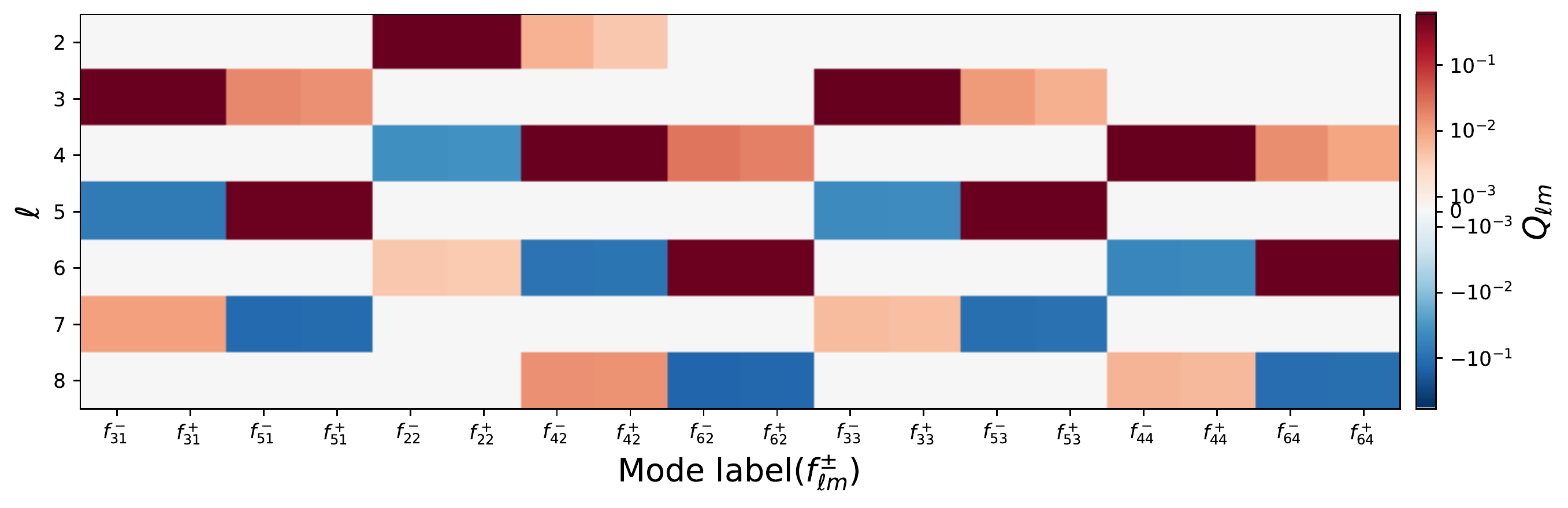}
    \includegraphics[width=\textwidth]{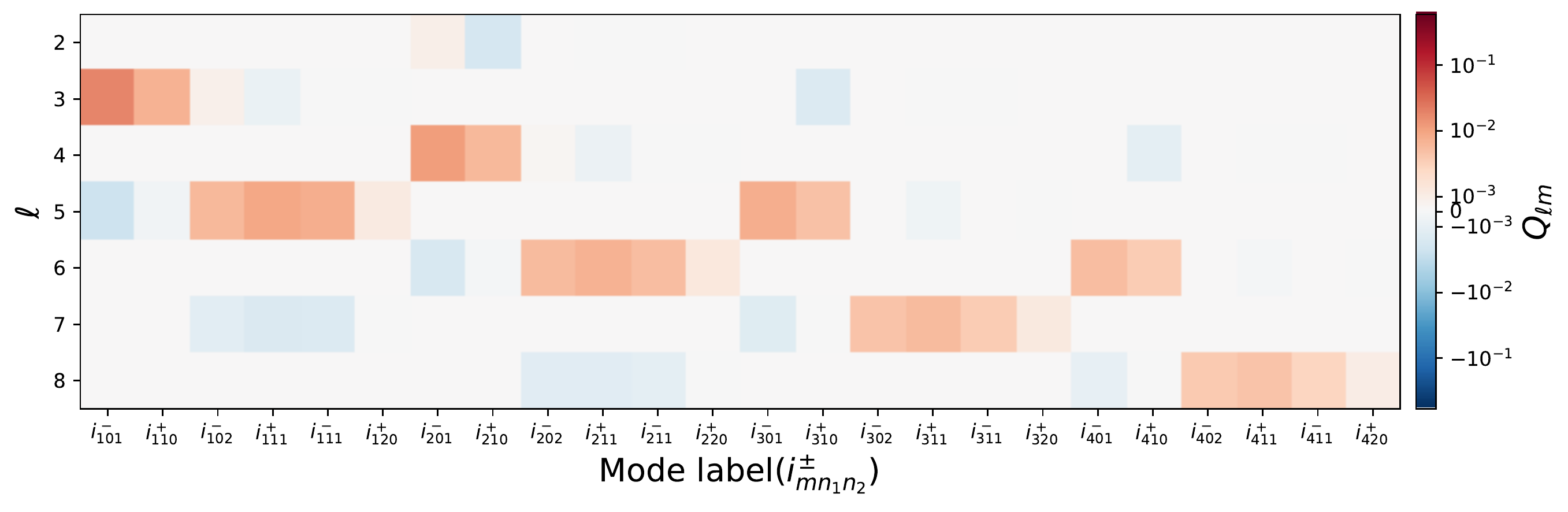}
    \caption{Colormaps showing, for each mode calculated from an $n=1$ polytrope with $\Omega/\Omega_d\simeq0.3$ (x-axis), values of $Q_{\ell m}$ for different $\ell$ (y-axis). Because of rotation, single modes can have non-zero $Q_{\ell m}$ for multiple $\ell$ (but only one $m,$ since the background state is still axisymmetric). }
    \label{fig:QvMode} 
\end{figure*}

\autoref{fig:QvMode} elaborates on \autoref{fig:n1pOm3Q}, providing a more complete breakdown of tidal coupling coefficients for different f-modes (top), and i-modes (bottom) calculated for an $n=1,$ $\Omega/\Omega_d\simeq0.3$ polytrope. For each mode labelled on the x-axis, the colormaps show $Q_{\ell m}$ for different spherical harmonic degrees $\ell$ (and the single $m$ associated with each mode). The color-plots for f-modes (top), and i-modes (bottom) are all saturated identically, and transition from log to linear scale near zero. 

Fundamental modes produce the largest $Q_{\ell m}$ by far. Inertial modes' $Q_{\ell m}$ values are smaller, but the frequencies of these oscillations generally lie much closer to resonance with the tidal forcing. Inertial modes also have a much stronger tesseral than sectoral overlap (e.g., the $m=2$ i-modes $i_{201}^-$ and $i_{210}^+$ have much larger $Q_{42}$ than $Q_{22}$). Additionally, with non-zero rotation the f-modes $f_{\ell m}^\pm$ invariably gain overlap coefficients $Q_{\ell+2,m}$ with the opposite sign of $Q_{\ell m}$. As described in Section \ref{sec:fmneg}, this sign difference produces the cross-correlations $Q_{\ell+2,m}Q_{\ell m}<0$ that are responsible for the divergence of tesseral Love numbers with increasing satellite separation (decreasing orbital frequency).

\end{document}